%% file: itw25_CQADC.tex
\documentclass[conference,letterpaper]{IEEEtran}
\IEEEoverridecommandlockouts

\def\Figs{./figs/} 

\usepackage[utf8]{inputenc} 
\usepackage[T1]{fontenc}
\usepackage{url}
\usepackage{ifthen}
\usepackage[cmex10]{amsmath} 

\newboolean{JOURNAL} 
\setboolean{JOURNAL}{false} 

\newboolean{TWOCOLUMN} 
\setboolean{TWOCOLUMN}{true} 

\newboolean{NOTE} 
\setboolean{NOTE}{false} 

\newboolean{arXiv} 
\setboolean{arXiv}{false} 

\newboolean{ITW_FINAL} 
\setboolean{ITW_FINAL}{true} 

\usepackage{mleftright}       
\mleftright                   
\usepackage{balance}
\usepackage{graphicx}         
\usepackage{booktabs}         

\usepackage{comment}
\usepackage[url,hyperrefblack,notheorems,IEEEtran]{research17} 
\usepackage{amsmath,amssymb,amsfonts,amsbsy}
\usepackage[colorinlistoftodos, textsize=footnotesize]{todonotes} 

\usepackage{amsthm} 
\newtheorem{theorem}{\mytheoremname}

\newtheorem{definition}{\mydefinitionname}
\newtheorem{remark}{\myremarkname}

\usepackage[font=footnotesize,,labelsep=period]{caption}
\usepackage{subcaption} 
\usepackage[capitalize]{cleveref}
\usepackage{tikz}
\usetikzlibrary{spy} 
\usepackage{pgfplots}
\usepackage[
  top=0.700in,
  bottom=1.080in,
  left=0.673in,
  right=0.673in
]{geometry}
\allowdisplaybreaks

\newcommand{\ADC}[1]{\textnormal{ADC}\left(#1\right)} 
\newcommand{\BSC}[1]{\textnormal{BSC}\left(#1\right)} 
\newcommand{\QSC}[1]{\textnormal{QSC}\left(#1\right)} 

\allowdisplaybreaks

\usepackage{array}
\usepackage{multirow}
\usepackage{enumitem} 
\usepackage{nicematrix} 
\newcommand*{\Scale}[2][4]{\scalebox{#1}{\ensuremath{#2}}} 
\usepackage{soul} 
\usepackage[colorinlistoftodos, textsize=footnotesize]{todonotes}
\newcommand{\hy}{\color{magenta}} 

\usepackage{commands}

\begin{document}
\title{On Finite-Blocklength Noisy Classical-Quantum Channel Coding With Amplitude Damping Errors} 


\ifthenelse{\boolean{JOURNAL}}{\author{Tam{\'a}s~Havas,~\IEEEmembership{Student Member,~IEEE}, Hsuan-Yin~Lin,~\IEEEmembership{Senior Member,~IEEE}, Eirik~Rosnes,~\IEEEmembership{Senior Member,~IEEE}, and Ching-Yi~Lai~\IEEEmembership{Senior Member,~IEEE}
\thanks{Tam{\'a}s~Havas, Hsuan-Yin~Lin, and Eirik~Rosnes are with Simula UiB, N--5006 Bergen, Norway. Email: \{tamas, lin, eirikrosnes\}@simula.no.}
\thanks{Ching-Yi~Lai is with the Institute of Communications Engineering, National Yang Ming Chiao Tung University, Taiwan.  Email: cylai@nycu.edu.tw.}}
}
{\author{%
   \IEEEauthorblockN{Tam{\'a}s~Havas, Hsuan-Yin~Lin, and Eirik~Rosnes}
     \IEEEauthorblockA{Simula UiB, 
       N--5006 Bergen, Norway\\
       Emails:  \{tamas, lin, eirikrosnes\}{@}simula.no}
     \and
     \IEEEauthorblockN{Ching-Yi Lai}
     \IEEEauthorblockA{Institute of Communications Engineering,\\ National Yang Ming Chiao Tung University,\\ 
       Hsinchu 300093, Taiwan, cylai@nycu.edu.tw}}
}

\maketitle


\begin{abstract}
  We investigate practical finite-blocklength \emph{classical–quantum channel coding} over the quantum amplitude damping channel (ADC), aiming to transmit classical information reliably through quantum outputs. Our findings indicate that for any finite blocklength, a naive (uncoded) approach fails to offer any advantage over the ADC. Instead, sophisticated encoding strategies that leverage both classical error-correcting codes and quantum input states are crucial for realizing quantum performance gains at finite blocklengths.
\end{abstract}

\section{Introduction}
\label{sec:introduction}

In quantum information theory, an important problem is to characterize the ultimate limits of reliably transmitting classical information over noisy quantum channels~\cite{Wilde17_1, Holevo19_1}. 

The \emph{classical capacity of a quantum channel} is defined as the maximum achievable transmission rate 
at which classical messages can be transmitted with arbitrarily small error probability, 
as the number of qubits used for encoding the classical messages (referred to as the \emph{blocklength} or the \emph{code length})  goes to infinity.   This capacity was characterized by the Holevo–Schumacher–Westmoreland (HSW) coding theorem, which extends Shannon’s foundational work in classical information theory to the quantum setting. The classical capacity is quantified by the Holevo information~\cite{Holevo98_1, SchumacherWestmoreland97_1}.

In recent decades, there has been an emerging trend of developing efficient coding schemes with \emph{finite} blocklength for various practical applications; see, e.g.,~\cite{PolyanskiyPoorVerdu10_1, Dalai13_1, AltugWagner14_1, MatthewsWehner14_1, Vazquez-VilarTausteCampoGuillenFabregasMartinez16_1, DurisiKochPopovski16_1, Polyanskiy17_1, WangFangTomamichel19_1, PanteleevKalachev21_1, MahmoodAtzeniJorswieckLopez23_1}. In the quantum setting, resource limitations are especially significant in the near-term, realistic noisy intermediate-scale quantum  era~\cite{Preskill18_1},
making large blocklengths impractical. 
Recently, there has been a renewed interest in the theoretical understanding of quantum communication theory with finite resources~\cite{Tomamichel16_1, TomamichelBertaRenes16_1, BlascoCollVazquez-VilarFonollosa22_1}, and most of the previous works focus on second-order asymptotics and moderate deviation analysis for classical communication over quantum noisy channels~\cite{TomamichelTan15_1, ChubbTanTomamichel17_1}. 
  These studies commonly rely on random coding arguments or conventional converse bounds to establish fundamental limits on the maximum achievable transmission rate under constrained resources and a fixed, sufficiently small error probability.
  However, the design of explicit, high-performance coding schemes for a given number of messages and an arbitrary (often short) blocklength remains largely an open and challenging problem.

In classical finite-blocklength information theory, several tight bounds on the best possible theoretical performance for a fixed sufficiently low error probability and a given blocklength have been derived since the influential work in~\cite{PolyanskiyPoorVerdu10_1}. Moreover, the design of good practical short-blocklength codes has attracted significant attention in recent years~\cite{ChenLinMoser13_1, IngberZamirFeder13_1, LinMoserChen18_1, CoskunDurisi-etal19_1, Arikan19_1sub}. Because a clear connection exists between the coding theorems for data transmission over classical noisy channels and those for transmitting classical information over quantum noisy channels, an important question arises: how can we leverage classical error-correcting codes to develop finite-blocklength coding schemes that approach the fundamental limits of classical communication over quantum channels?

This work focuses on the performance analysis of finite-blocklength channel coding for \ifthenelse{\boolean{JOURNAL}}{both classical communication and \emph{entanglement-assisted (EA)} classical communication}{classical communication} over a noisy classical-quantum (CQ) channel, formulated as a noiseless CQ channel followed by a qubit \emph{amplitude damping channel (ADC)}. We quantify the performance of a coding scheme using the \emph{(average) probability of successful decoding} for classical information, also referred to as the \emph{logical error rate}. In this framework---often termed \emph{CQ channel coding}~\cite{Wilde17_1, Holevo19_1}---a coding scheme is generally described as an $(n,\const{M})_{q}$ code, along with a given associated set of quantum input states, where $n$ is the code length, $\const{M}$ denotes the number of codewords, and $q$ is the field size, or as an $[n,k]_{q}$ code, with $k$ being the code dimension.
 
Two main CQ channel coding strategies are considered for classical communication over a noisy CQ channel, without the assistance of shared entanglement at the receiver end.
The first, simpler approach involves performing \emph{individual measurements} on the tensor-product outputs of the encoding states at the receiver. This method effectively induces a purely classical discrete memoryless (DM) channel, simplifying the analysis of classical information transmission over a quantum channel. The second, more complex approach involves performing a \emph{collective measurement} on the output states at the receiver. While this approach also induces a purely classical channel, the resulting channel is generally more difficult to characterize. Analyzing the coding performance in this scenario typically requires advanced numerical methods, such as semidefinite programming (SDP).

Our main contributions are listed as follows.
\begin{enumerate}[nosep,label=\roman*)]
\item For a single use of an ADC (i.e., blockelength $n=1$), we present the optimal classical code, input states, and the corresponding optimal measurements employed at the receiver end in terms of average success probability. It is shown that the optimal input states to achieve the optimal average success probability are different from the best input states that achieve the single-letter \emph{Holevo $\chi$-capacity}~\cite[Def.~13.3.1]{Wilde17_1} for the ADC (see Theorem~\ref{thm:optimal-rho_cqADC_single-use}).

\item We compare the finite-blocklength performance of the \emph{individual-measurements} strategy with that of the \emph{collective-measurement} strategy for the ADC. Theorem~\ref{thm:uncoding_not-help_ADC} indicates that for uncoded transmission (i.e., $n=k$), the receiver does not need to perform a collective measurement on multiple channel outputs to improve performance\textemdash performing individual optimal measurements for the single-use ADC at the receiving end is sufficient. Nevertheless, we demonstrate that the collective-measurement strategy can be beneficial compared to the individual-measurements strategy when an $[n,k]_{q}$ code with $n>k$ is employed (see Section~\ref{sec:motivating-example}).

\ifthenelse{\boolean{JOURNAL}}{
\item For EA classical communication over the ADC, we present a similar result to Theorem~\ref{thm:optimal-rho_cqADC_single-use} for the single-use EA ADC. We show that Bell states outperform all other pure input states in terms of average success probability, and we identify the corresponding optimal measurements to be employed at the receiver end (see Theorem~\ref{thm:best-pure-rho_EA-cqADC_single-use}).

\item Since employing the individual-measurements strategy for the EA ADC induces a classical DM $4$-ary asymmetric channel ($4$-AC), we derive new finite-blocklength upper and lower bounds on the success probability over the $q$-AC for any positive integer $q$ to compare the performance of the individual-measurements strategy with that of the collective-measurement strategy (see Theorems~\ref{thm:UB_qAC} and~\ref{thm:LB_qAC}).}
{}

\item Finally, in Section~\ref{sec:numerical-results}, we present numerical results comparing the finite-blocklength performance of coding schemes employing the individual-measurements strategy with those using the collective-measurement strategy for \ifthenelse{\boolean{JOURNAL}}{both the ADC and the EA ADC.}{the ADC.} Notably, the collective-measurement strategy consistently achieves strictly better performance than that of the individual-measurements strategy.
\end{enumerate}

\ifthenelse{\boolean{ITW_FINAL}}{All proofs are omitted due to space constraints.}
{
  \ifthenelse{\boolean{arXiv}}{}{Some proofs are omitted due to space constraints and can be found in~\cite{HavasLinRosnesLai25_1sub}.}
}

\section{Preliminaries and Setup}
\label{sec:preliminaries-setup}

\subsection{Notation}
\label{sec:notation}

We denote the set of natural numbers by $\Naturals$. For $i,j \in\Naturals\cup\{0\}$ with $i \leq j$, let $[i:j] \eqdef \{i, i+1, \ldots, j\}$. \emph{Row} vectors are represented by boldfaced letters, e.g., $\vect{x}$. Matrices and sets are denoted by capital sans-serif letters and calligraphic uppercase letters, respectively. For example, $\mat{X}$ denotes a matrix and $\set{X}$ denotes a set. 
A classical code $\code{C}$ over a finite field $\Field_q$ (of size $q$) is a subset of $\Field_q^n$ with codewords. A code with $\const{M}$ codewords is an $(n, \const{M})_q$ code, and if the minimum distance between distinct codewords is $d$, it is an $(n, \const{M}, d)_q$ code. A linear code of dimension $k$ is an $[n, k]_q$ or $[n, k, d]_q$ code.
The set of linear operators on a Hilbert space $\set{H}$ is denoted by $\collect{L}(\set{H})$, and the set of density operators by $\collect{D}(\set{H}) \subset \collect{L}(\set{H})$. We use standard Dirac notation: a pure quantum state can be represented by a column vector $\ket{\psi} \in \set{H}$, while a general quantum state is represented by a density operator $\rho\in \collect{D}(\set{H})$. The column vectors $\ket{1},\ldots,\ket{n}$ represent the standard basis of an $n$-dimensional Hilbert space. The tensor product of operators $\mat{A}_1, \ldots, \mat{A}_n$ is denoted by $\bigotimes_{j=1}^n \mat{A}_j \eqdef \mat{A}_1 \otimes \cdots \otimes \mat{A}_n$, and $\mat{A}^{\otimes n}\eqdef\mat{A}\otimes\cdots\otimes\mat{A}$ ($n$ terms).

\subsection{Channel Coding for Classical Communication Over  Noisy Classical-Quantum Channels}
\label{sec:channel-coding_classical_quantum-channel}

Consider the transmission of $\const{M} \in \Naturals$ classical messages over multiple uses of a noisy CQ channel. 
This noisy CQ channel can be modeled as a composition of a noiseless CQ map $\code{S} = \{\rho_x\}_{x \in \mathcal{X}} \subset \collect{D}(\set{H}^{\otimes n})$, which encodes information into quantum states, followed by a noisy quantum channel $\set{N}$. We denote this combined channel as $(\code{S}, \set{N})$. Specifically, the transmission process is described by 
\begin{equation*} 
 \mathcal{X} \stackrel{\code{S}}{\longmapsto} \{\rho_x\}_{x \in \mathcal{X}} \stackrel{\set{N}}{\longmapsto} \{\set{N}(\rho_x)\}_{x \in \mathcal{X}}.
\end{equation*}

\begin{definition}[{An $(n,\const{M})_{q}$ Code}]
  \label{def:cq-code_Mn}
  An $(n,\const{M})_{q}$ coding scheme for a noisy CQ channel $(\code{S}=\{\rho_{x}\}_{x\in\set{X}}, \set{N})$ over an input alphabet $\set{X}$ of size $\absol{\set{X}} = q$ with $n$ channel uses consists of a set of $\const{M}$ length-$n$ codewords $\code{C}=\{\vect{x}_1,\ldots,\vect{x}_\const{M}\}$ 
  such that for  $m\in [1:\const{M}]$,\footnote{Note that there is a one-to-one mapping of the field $\mathbb{F}_q$ to the channel input alphabet $\mathcal{X}$.}
  \begin{IEEEeqnarray*}{rCl}
    \IEEEeqnarraymulticol{3}{l}{%
      \vect{x}_m=(x_{m,1},\ldots,x_{m,n})\in\set{X}^n}\nonumber\\*\hspace*{5mm}%
    & \mapsto &\set{N}^{\otimes n}(\rho_{\vect{x}_m})\eqdef\set{N}(\rho_{x_{m,1}})\otimes\cdots\otimes\set{N}(\rho_{x_{m,n}}).\IEEEeqnarraynumspace\label{eq:encoding_cq-code}
  \end{IEEEeqnarray*}

\end{definition}

To decode the output state, a quantum measurement is performed on $\set{H}^{\otimes n}$, which is described by a \textit{positive operator-valued measure} (POVM) with a set of nonnegative Hermitian operators, $\set{D}\eqdef\{\Lambda_m\}_{m\in [1:\const{M}]}\subseteq\collect{L}(\set{H})$ satisfying $\sum_{m \in [1:\const{M}]} \Lambda_m = \mat{I}$, where $\mat{I}$ is the identity matrix.

Throughout this paper, we assume that the messages are uniformly distributed. The optimal POVM that maximizes the \emph{average success probability} for a given $(n,\const{M})_{q}$ code can be determined through \emph{quantum hypothesis testing}~\cite[Sec.~II]{BlascoCollVazquez-VilarFonollosa22_1}.

\begin{definition}[An Optimal $(n,\const{M})_{q}$ Code]
  \label{def:optimal-POVM}
  Consider a noisy CQ channel  ($\code{S}=\{\rho_{x}\}_{x\in\set{X}}, \set{N})$  and an $(n,\const{M})_{q}$ code $\code{C}$.
  A POVM $\set{D}^\star\eqdef\{\Lambda^\star_1,\ldots,\Lambda^\star_\const{M}\}$ is called optimal if
  \begin{IEEEeqnarray}{c}
    \set{D}^\star=\argmax_{\set{D}}\{P_{\textnormal{c}}(\code{C},\code{S},{\cal N},\set{D})\},
    \label{eq:maximum_average-success-prob}
  \end{IEEEeqnarray}
  where 
  the average success probability is given by
  \begin{equation*}
    P_{\textnormal{c}}(\code{C},\code{S},{\cal N},\set{D})\eqdef\frac{1}{\const{M}}\sum_{m\in[1:\const{M}]}\bigtrace{\Lambda_m{\set{N}}^{\otimes n}(\rho_{\vect{x}_m})}.
  \end{equation*}
  Define $P_{\textnormal{c}}(\code{C},\code{S})\eqdef P_{\textnormal{c}}(\code{C},\code{S}, \set{N},\set{D}^\star)$. An $(n,\const{M})_{q}$ code $\code{C}^\star$ with an associated $\code{S}^\star$   is called optimal if
  \begin{IEEEeqnarray*}{c}
    P^\star_{\textnormal{c}}\eqdef P_{\textnormal{c}}(\code{C}^\star,\code{S}^\star)\geq P_{\textnormal{c}}(\code{C},\code{S})
  \end{IEEEeqnarray*}
  for any $\code{C}$ and $\code{S}$.

  Furthermore, when the associated $\code{S}$ is fixed, we will denote the average success probability simply as $P_{\textnormal{c}}(\code{C})\triangleq P_{\textnormal{c}}(\code{C}, \code{S})$.
\end{definition}
The code $\code{C}$ will be denoted as $\code{C}^{[n,k]_q}$, $\code{C}^{(n,\const{M})_q}$, or $\mat{G}_q$ to indicate that it is a linear code, a nonlinear code, or a code defined by a generator matrix $\mat{G}_q$ over the finite field $\Field_q$, respectively. With slight abuse of notation, when the input alphabet $\set{X}=\{0,1\}$ is clear from the context, the subscript indicating the field size will be omitted.

For small dimensions, \eqref{eq:maximum_average-success-prob} can be solved numerically via  SDP, provided that the states in $\code{S}$ are given.

\subsubsection{The Amplitude Damping Channel}
The ADC is a noisy quantum channel $\mathcal{N}_\gamma$  where 
the channel output on an input state $\rho \in \{\rho_0,\rho_1\}$ $\left(\set{X} = \{0,1\}\right)$ is
\begin{IEEEeqnarray*}{c}
  \set{N}_\gamma(\rho)=\sum_{i\in\{0,1\}}\mat{K}_i\rho\hermi{\mat{K}_i},
\end{IEEEeqnarray*}
where $\hermi{(\cdot)}$ is the adjoint of a matrix, $\mat{K}_0=\bigl[
\begin{smallmatrix}
  1 & 0
  \\
  0 & \sqrt{1-\gamma}
\end{smallmatrix}\bigr]$ and 
$\mat{K}_1=\bigl[
\begin{smallmatrix}
  0 & \sqrt{\gamma}
  \\
  0 & 0
\end{smallmatrix}\bigr]$ 
are its \emph{Kraus operators}, and $\gamma$ is the damping parameter with $0\leq\gamma\leq 1$.

\subsection{The Induced Classical Channel}
\label{sec:induction-purely-classical-channel}

Consider the $(n,\const{M})_{q}$ coding scheme described in Definition~\ref{def:cq-code_Mn}. If the receiver performs \emph{individual measurements} on the tensor-product output state $\set{N}^{\otimes n}(\rho_{\vect{x}})$ using a POVM $\{\Lambda_y\}_{y\in\set{Y}}$  for some output alphabet $\set{Y}$, this approach induces the  conditional probability distribution
\begin{IEEEeqnarray*}{c}
  P_{\vect{Y}|\vect{X}}(\vect{y}|\vect{x})=\prod_{j=1}^n\bigtrace{\Lambda_{y_j}\set{N}(\rho_{x_{j}})},
\end{IEEEeqnarray*}
which is characterized by an \emph{independent and identically distributed} structure across channel uses. Consequently, $P_{Y|X}(y|x)\triangleq\trace{\Lambda_y\set{N}(\rho_x)}$ serves as the \emph{channel law} of an equivalent classical DM channel.

On the other hand, for a finite blocklength $n$, the receiver may perform a \emph{collective measurement} on the entire tensor-product output state using a POVM $\set{D} = \{\Lambda_m\}_{m \in [1:\const{M}]}$. This induces the  conditional probability distribution
 \begin{IEEEeqnarray*}{c}
  P_{M|\vect{X}}(m|\vect{x})=\bigtrace{\Lambda_{m}\set{N}^{\otimes n}(\rho_{\vect{x}})}.
\end{IEEEeqnarray*}
We will primarily compare the performance of decoding strategies based on individual measurements of tensor-product output states with those based on collective measurements.

\ifthenelse{\boolean{JOURNAL}}{
\subsection{Channel Coding for Entanglement-Assisted (EA) Classical Communication Over a Classical-Quantum Channel}
\label{sec:channel-coding_EA-classical-communication_quantum-channel}

In the quantum communication literature, it is demonstrated that the rate of classical data transmission over a quantum channel can be improved when infinite prior entanglement resources are available~\cite[Ch.~21]{Wilde17_1}. Similar to the channel coding scenario of classical communication over a noisy  CQ channel ($\code{S}, \set{N}$), this work specifically focuses on a channel coding scheme inspired by \emph{super-dense coding}~\cite[Ch.~6]{Wilde17_1} for EA classical communication over a quantum channel. Consequently, an EA $(n,\const{M})$ code can be defined similarly to Definition~\ref{def:cq-code_Mn}, except that the code symbol is defined over an alphabet $\set{X}^2$ and the associated set of input states $\{\rho_{\vect{x}_m}\}_{m\in [1:\const{M}]}$ is chosen such that
\begin{IEEEeqnarray*}{rCl}
  \IEEEeqnarraymulticol{3}{l}{%
    \vect{z}_m=(z_{m,1},\ldots,z_{m,n})\in(\set{X}^2)^n}\nonumber\\*\hspace*{1mm}%
  & \mapsto &\rho_{\vect{z}_m}=\rho_{z_{m,1}}\otimes\cdots\otimes\rho_{z_{m,n}}\in\set{D}((\set{H}^2)^{\otimes n}),\, m\in [1:\const{M}].\IEEEeqnarraynumspace\label{eq:encoding_cq-EAcode}
\end{IEEEeqnarray*}
Similar to Definition~\ref{def:optimal-POVM}, it can be shown that the success probability of an EA $(n,M)$ code can be determined by performing a POVM on the tensor-product output states $(\set{N}\otimes\mat{I})^{\otimes n}(\rho_{\vect{z}_m})$.

The \emph{EA classical capacity} of a quantum channel is the maximum amount of classical information that can be transmitted when unlimited entanglement is shared between the transmitter and receiver~\cite{BennettShorSmolinThapliyal99_1}. It is known that the EA classical capacity of the ADC is equal to~\cite{GiovannettiFazio05_1}
\begin{IEEEeqnarray*}{c}
  \const{C}_{\textnormal{EA}}(\set{N}_\gamma)=\max_{p\in [0,1]}\{\Hb(p)+\Hb((1-\gamma)p)-\Hb(\gamma p)\}.
\end{IEEEeqnarray*}
\todo[author=Lin,inline]{This is a concise version of EA-CQ coding; it may require updates at a later stage.}
}{}

\subsection{Channel Capacity}
\label{sec:capacities-channel-model}

The maximum achievable rate for transmitting classical information over a noisy quantum channel $\set{N}$ with input alphabet $\set{X}$ is defined as the channel’s \emph{classical capacity}~\cite[Th.~20.3.1]{Wilde17_1}, \cite{Holevo98_1, SchumacherWestmoreland97_1}. A   lower bound on this capacity is given by the \emph{Holevo $\chi$-capacity}, defined as
\begin{IEEEeqnarray*}{c}
  \chi(\set{N})\eqdef{\max_{\{p_{x},\rho_x\}_{x\in\set{X}}}\Scale[0.9]{\biggl\{\biggHP{\sum\limits_{x\in\set{X}}p_x\set{N}(\rho_x)}-\sum\limits_{x \in\set{X}}p_x\bigHP{\set{N}(\rho_x)}\biggr\}}},
\end{IEEEeqnarray*}
where the maximization is taken over all input distributions $p_x$ and states $\rho_x$ for $x \in \set{X}$, and $\eHP{\cdot}$ denotes the \emph{von Neumann entropy}.

In this work, we focus on the transmission of classical information over the qubit ADC $\set{N}_\gamma$ with input alphabet $\set{X}=\{0,1\}$, for which the classical capacity remains unknown~\cite[Sec.~4.7.5]{Wilde17_1}. However, it is known that its Holevo $\chi$-capacity is given by~\cite{GiovannettiFazio05_1}
\begin{IEEEeqnarray}{rCl}
  \chi(\set{N}_\gamma)& = &\max_{p\in [0,1]}\Biggl\{\Hb((1-\gamma)p)\nonumber\\
  &&\hspace*{1.25cm} -\>\Hb\Biggl(\frac{1+\sqrt{1-4(1-\gamma)\gamma p^2}}{2}\Biggr)\Biggr\},
  \IEEEeqnarraynumspace\label{eq:chi-capacity_ADC}
\end{IEEEeqnarray}
where $\Hb(\cdot)$ is the binary entropy function. The $\chi$-capacity-achieving input distribution and states are given by $p_0 = p_1 = \nicefrac{1}{2}$ and 
\begin{IEEEeqnarray}{c}
  \code{S}_{\chi}\eqdef\left\{
    \rho_{\pm}\eqdef\begin{pmatrix}
      1-p^\star & \pm\sqrt{(1-p^\star)p^\star}
      \\[1mm]
      \pm\sqrt{(1-p^\star)} & p^\star
    \end{pmatrix}\right\},\label{eq:opt_states_chi-capacity}
\end{IEEEeqnarray}
respectively, where $p^\star$ is the solution that maximizes~\eqref{eq:chi-capacity_ADC}.

\section{Performance Analysis for Classical Communication Over the ADC}
\label{sec:performance-analysis_CQADC}

In this section, we analyze the performance of channel coding for classical communication over an ADC. 

\subsection{Optimal Coding Over a Single-Use Qubit ADC}
\label{sec:optimal-code_ADCn1}

We begin by presenting our initial results through the characterization of the optimal $(n,\const{M}) = (1,2)_2$ code for the ADC $\set{N}_\gamma$. In the single-use setting, the only possible binary code is trivially given by $\code{C}^{(1,2)} = \{0,1\}$. The remaining task is to determine the corresponding optimal input states.

\begin{theorem}
  \label{thm:optimal-rho_cqADC_single-use}
  For an ADC $\set{N}_{\gamma}$ with damping parameter $0\leq\gamma\leq 1$, the optimal $(n,\const{M})=(1,2)_2$ code is $\code{C}^{\star}=\{0,1\}$ with its associated $\code{S}^{\star}=\code{S}^\star_{\textnormal{pm}}\eqdef\{\rho_{0}=\ketbra{+}{+}, \rho_1=\ketbra{-}{-}\}$, where $\ket{+}\eqdef\nicefrac{(\ket{0}+\ket{1})}{\sqrt{2}}$ and $\ket{-}\eqdef\nicefrac{(\ket{0}-\ket{1})}{\sqrt{2}}$. Moreover, the optimal POVM is $\set{D}^\star=\set{D}^\star_{\textnormal{pm}}\eqdef\{\Lambda_0=\rho_0, \Lambda_1=\rho_1\}$ and the induced channel is a classical binary symmetric channel (BSC) with crossover probability $\eps_{\textnormal{BSC}}=\nicefrac{(1-\sqrt{1-\gamma})}{2}$.
\end{theorem}
\ifthenelse{\boolean{ITW_FINAL}}{}{
\begin{IEEEproof}
  See \ifthenelse{\boolean{arXiv}}{Appendix~\ref{sec:proof_Theorem1}.}{\cite[App.~A]{HavasLinRosnesLai25_1sub}.}
\end{IEEEproof}}

\begin{remark}
  \label{rem:remark1} 
  We remark that the optimal input states $\code{S}^{\star}_\textnormal{pm}$ for the single-use ADC differ from the $\chi$-capacity-achieving states $\code{S}_{\chi}$ given in~\eqref{eq:opt_states_chi-capacity}. Using $\code{S}_{\chi}$ leads to a strictly sub-optimal performance, i.e., $P_{\textnormal{c}}(\code{C}^\star,\code{S}_{\chi})<P_{\textnormal{c}}(\code{C}^\star,\code{S}_{\textnormal{pm}}^\star)$.
\end{remark}

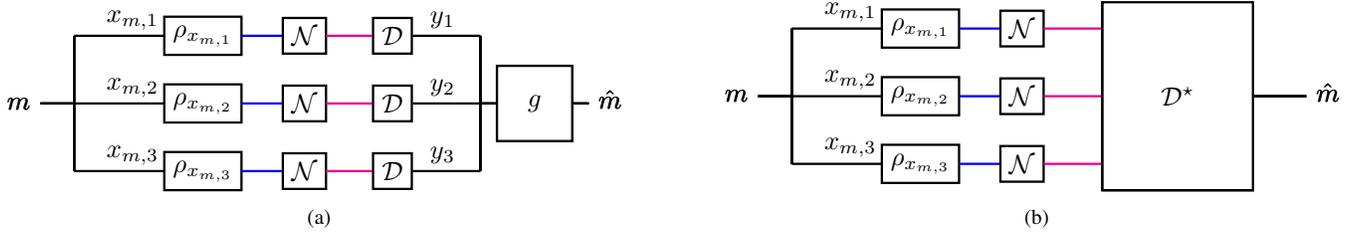
\begin{figure*}[t!]
  \centering
  \begin{subfigure}{0.475\linewidth}
    \centering
    \input{\Figs/coding_cqadc_M2n3_indep.tex}
    \subcaption{}
    \label{fig:coding_cqadc_M2n3_indep}
  \end{subfigure}
  \hfill
  \begin{subfigure}{0.475\linewidth}
    \centering
    \input{\Figs/coding_cqadc_M2n3_joint_outputs.tex}
    \subcaption{}
    \label{fig:coding_cqadc_M2n3_joint}
  \end{subfigure}  
  \caption{Comparison of \subref{fig:coding_cqadc_M2n3_indep} individual measurements and \subref{fig:coding_cqadc_M2n3_joint} a collective measurement on the output states.
 }
\end{figure*}

\subsection{Uncoded Strategies Do Not Benefit From Collective Measurements}
\label{sec:uncoding_not-help}

In this subsection, we show that extending the uncoded approach to multiple channel uses offers no performance gain from using collective measurements. Specifically, performing a collective measurement on uncoded transmissions yields the same average success probability as individual measurements.

\begin{theorem}
  \label{thm:uncoding_not-help_ADC}
  Consider an ADC $\set{N}_\gamma$ with damping parameter $0\leq\gamma\leq 1$ and a \emph{trivial} code $\code{C}^{[n,n]}=\{0,1\}^n$ with a given associated $\code{S}^\star_\textnormal{pm}=\{\rho_0=\ketbra{+}{+}, \rho_1=\ketbra{-}{-}\}$, where $n\geq 1$. 
  Suppose the receiver performs individual measurements using the optimal single-use ADC POVM $\set{D}^\star_\textnormal{pm} = {\Lambda_0 = \ketbra{+}{+},\ \Lambda_1 = \ketbra{-}{-}}$ as specified in Theorem~\ref{thm:optimal-rho_cqADC_single-use}. Then, the average success probability of $\code{C}^{[n,n]}$ under this strategy is equal to that achieved by performing an optimal collective measurement, i.e.,
  \begin{IEEEeqnarray*}{c}    P_\textnormal{c}\bigl(\code{C}^{[n,n]},\set{D}^{\textnormal{ML}}\bigr)=P_\textnormal{c}(\code{C}^{[n,n]}).
  \end{IEEEeqnarray*}
  Here, we use the classical maximum likelihood (ML) decoding approach to define $\set{D}^{\textnormal{ML}}\eqdef\{\Lambda^{\textnormal{ML}}_m\}_{m\in [1:\const{M}]}$ where
  \begin{IEEEeqnarray*}{c} \Lambda^{\textnormal{ML}}_m\eqdef\sum_{\substack{\vect{y}:\\g(\vect{y})=m}}\bigotimes_{j\in [1:n]}\Lambda_{y_j},\, m\in [1:\const{M}],
  \end{IEEEeqnarray*}
$y_j\in\{0,1\}$, $\Lambda_{y_j}\in\set{D}^\star_{\textnormal{pm}}$, $j\in [1:n]$, and
  \begin{IEEEeqnarray*}{c}
    g(\vect{y})\eqdef\argmax_{m\in [1:\const{M}]}\prod_{j=1}^n\bigtrace{\Lambda_{y_j}\set{N}_\gamma(\rho_{x_{m,j}})}.
  \end{IEEEeqnarray*}
\end{theorem}
\ifthenelse{\boolean{ITW_FINAL}}{}
{
  \ifthenelse{\boolean{arXiv}}{
    \begin{IEEEproof}
      See Appendix~\ref{sec:proof_uncoding_not-help_EAcqADC}.
    \end{IEEEproof}}
}
\ifthenelse{\boolean{ITW_FINAL}}{}{The detailed proof can be found in the extended version~\cite[App.~B]{HavasLinRosnesLai25_1sub}.} Next, we sketch the main techniques of the proof. First, using the definition of $\set{D}^{\textnormal{ML}}$, it can be seen that for the uncoded approach, $\set{D}^{\textnormal{ML}}$ is $\bigl\{\bigotimes_{j\in [1:n]}\Lambda_{x_{m,j}}\bigr\}_{m\in[1:\const{M}]}$, where $\vect{x}_m\in\code{C}^{[n,n]}$ is the $m$-th codeword. Thus, we only need to identify that the optimal POVM $\set{D}^\star$ for the uncoded approach is identical to $\bigl\{ \bigotimes_{j\in [1:n]}\Lambda_{x_{m,j}}\bigr\}_{m\in[1:\const{M}]}$. This can be done by verifying the Holevo-Yuen-Kennedy-Lax conditions~\cite[Lem.~3]{BlascoCollVazquez-VilarFonollosa22_1}, and we can show a sufficient condition\ifthenelse{\boolean{ITW_FINAL}}{}{~\cite[Lem.~5]{HavasLinRosnesLai25_1sub}} for such tensor-product POVMs to satisfy these conditions.

Theorem~\ref{thm:uncoding_not-help_ADC} implies that to benefit from the power of collective quantum measurements, one must first encode messages using a nontrivial code across multiple channel uses.

\subsection{Motivating Example: Transmitting Four Messages Over Three Channel Uses}
\label{sec:motivating-example}

Based on the observation from Theorem~\ref{thm:uncoding_not-help_ADC}, we seek to design a coding scheme with $n > k$ and investigate whether such a scheme can improve performance through collective measurements on the channel outputs, as opposed to individual measurements.
 In this subsection, we present a motivating example in which the sender aims to transmit four messages ($\const{M} = 4$) over three uses ($n = 3$) of the ADC $\set{N}_{\gamma}$. To this end, we employ a $[3,2]_2$ single parity-check (SPC) code $\code{C}^{[3,2]}_{\textnormal{ SPC}}$ with generator matrix
$\bigl[\begin{smallmatrix}
        1 & 0 & 1
        \\
        0 & 1 & 1
\end{smallmatrix}\bigr]$ and the associated $\code{S}^\star_{\textnormal{pm}}=\{\rho_0=\ketbra{+}{+}, \rho_1=\ketbra{-}{-}\}$ for encoding.

\subsubsection{Basic Scheme: Measuring Outputs Individually}
\label{sec:basic-scheme}

The basic approach is to perform individual measurements on the outputs of $\set{N}_{\gamma}^{\otimes 3}$, so that each channel use induces an effective  BSC channel:
\begin{IEEEeqnarray*}{c}
  \bigtrace{\Lambda_y\set{N}_\gamma(\rho_x)}=
  \begin{cases}
    1-\eps_\textnormal{BSC}, & \textnormal{if } x=y;
    \\
    \eps_\textnormal{BSC}, & \textnormal{if } x\neq y,
  \end{cases}\quad x,y\in \{0,1\},
\end{IEEEeqnarray*}
where the crossover probability $\eps_\textnormal{BSC}=\nicefrac{(1-\sqrt{1-\gamma})}{2}$ is given by Theorem~\ref{thm:optimal-rho_cqADC_single-use}. The basic approach is illustrated in Fig.~\ref{fig:coding_cqadc_M2n3_indep}. To correctly decode a received vector $\vect{y}=(y_1,y_2,y_3)\in\{0,1\}^{3}$, we use the classical ML decoder $g(\vect{y})$ as specified by Theorem~\ref{thm:uncoding_not-help_ADC}. Then, the average success probability of the code $\code{C}^{[3,2]}_\textnormal{SPC}$ using the POVM $\set{D}^{\textnormal{ML}}$ is given by
\begin{IEEEeqnarray*}{c}
  P_{\textnormal{c}}\bigl(\code{C}_\textnormal{SPC}^{[3,2]},\set{D}^{\textnormal{ML}}\bigr)
  =\frac{1}{\const{M}}\sum_{m\in [1:4]} \trace{\Lambda^{\textnormal{ML}}_{m}\set{N}_{\gamma}^{\otimes 3}(\rho_{\vect{x}_m})}.
\end{IEEEeqnarray*}

\subsubsection{Improved Scheme:  Collective Measurement on the Outputs}
\label{sec:improved-scheme}

We choose the same $[n,k]_2=[3,2]_2$ SPC code with 
\begin{IEEEeqnarray*}{c}
  \bigtrace{\Lambda^\star_{\hat{m}} \set{N}_\gamma^{\otimes 3}(\rho_{\vect{x}_m})}=
  \begin{cases}
    1-3\eps_\textnormal{QSC}, & \textnormal{if } m=\hat{m};
    \\
    \eps_\textnormal{QSC}, & \textnormal{if } m\neq \hat{m},
  \end{cases}
\end{IEEEeqnarray*}
for $m,\hat{m}\in [1:4]$, where $\set{D}^\star=\{\Lambda_1^\star,\ldots,\Lambda_4^\star\}$ is the optimal POVM defined in Definition~\ref{def:optimal-POVM}. As a result, this leads to an induced quaternary symmetric channel (QSC) with crossover probability $\eps_{\textnormal{QSC}}$. Moreover, $P_{\textnormal{c}}\bigl(\code{C}_\textnormal{SPC}^{[3,2]}, \set{D}^{\star}\bigr)$ is equal to the average success probability using ML decoding of a quaternary code with generator matrix $[1]_4$ on the QSC. The improved scheme is shown in Fig.~\ref{fig:coding_cqadc_M2n3_joint}.

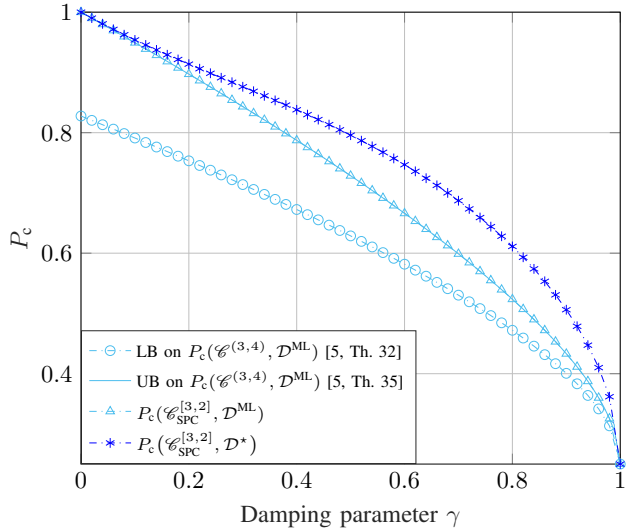
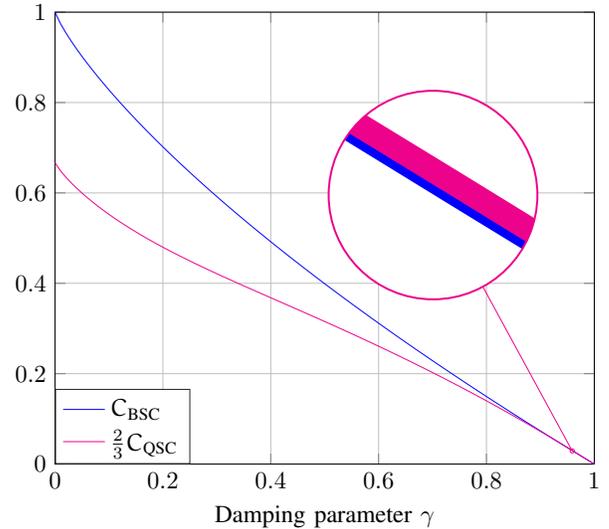
\begin{figure*}[t!]
  \centering
  \begin{subfigure}{0.475\linewidth}
    \Scale[0.925]{\input{\Figs/success_prob_versus_gamma_cqadc_M4n3_v1.tex}}
    \subcaption{}
    \label{fig:performance_bounds_ADCequivalent_qSC}
  \end{subfigure}
  \hfill
  \begin{subfigure}{0.475\linewidth}
    \centering
    \Scale[0.925]{\input{\Figs/CQADC_capacities.tex}}
    \subcaption{}
    \label{fig:capacities_ADCequivalent_qSCs}
  \end{subfigure}
  \caption{\subref{fig:performance_bounds_ADCequivalent_qSC} Performance bounds on the average success probability of the $[n,k]_2=[3,2]_2$ SPC code over the ADC $\set{N}_{\gamma}$ for   $0\leq\gamma\leq 1$. \subref{fig:capacities_ADCequivalent_qSCs} A comparison between the induced $\BSC{\eps_\textnormal{BSC}}$ and $\QSC{\eps_{\textnormal{QSC}}}$ from the ADC $\set{N}_{\gamma}$ for  $0\leq\gamma\leq 1$.}
  \vspace{-2ex}
\end{figure*}
Finally, we conclude this example by summarizing the average success probability in Fig.~\ref{fig:performance_bounds_ADCequivalent_qSC} from a classical channel coding perspective. The upper and lower bounds (UB and LB) on the average success probability of an arbitrary code $\code{C}^{(3,4)}$ over the $\BSC{\eps_\textnormal{BSC}}$ are equivalent to those on $P_\textnormal{c}\bigl(\code{C}^{(3,4)},\set{D}^{\textnormal{ML}}\bigr)$, and can be calculated exactly as shown in~\cite{PolyanskiyPoorVerdu10_1, ChenLinMoser13_1}. Note that in Fig.~\ref{fig:performance_bounds_ADCequivalent_qSC}   the best performance, the UB on $P_\textnormal{c}\bigl(\code{C}^{(3,4)},\set{D}^{\textnormal{ML}}\bigr)$, is achieved by $P_\textnormal{c}\bigl(\code{C}^{[3,2]}_\textnormal{SPC},\set{D}^{\textnormal{ML}}\bigr)$, indicating that when performing individual measurements, the SPC code is optimal. On the other hand, it also shows that the improved scheme using a collective measurement can be strictly better than the best individual measurement-based scheme, i.e., $P_\textnormal{c}(\code{C}^{[3,2]}_\textnormal{SPC},\set{D}^\star)>P_\textnormal{c}\bigl(\code{C}^{[3,2]}_\textnormal{SPC},\set{D}^{\textnormal{ML}}\bigr)$ for any $0\leq \gamma\leq 1$.

\subsection{Asymptotic Analysis}
\label{sec:asymptotic-analysis}

Continuing with Section~\ref{sec:motivating-example}, we provide a brief comparison of the best possible achievable rates between the basic and improved approaches, considering channel coding with a large blocklenth $n$ over the equivalent BSC and QSC,  i.e., the comparison of their capacities. 
For the basic scheme, Shannon's channel coding theorem states that there exists an $(n,\const{M})_2=(n,2^{n\const{C}_\textnormal{BSC}})_2$ coding scheme with an arbitrarily small error probability, where $\const{C}_\textnormal{BSC}=1-\Hb(\eps_\textnormal{BSC})$. Similarly, if we consider classical codes over $\Field_4$ for the equivalent $\QSC{\eps_\textnormal{QSC}}$, one can also conclude that an $(n',\const{M}')_4=(n',4^{n'\const{C}_\textnormal{QSC}})_4$ good code exists with an arbitrarily small error probability, where
\begin{IEEEeqnarray*}{c}
  \const{C}_\textnormal{QSC}\eqdef\frac{2-\Hb(3\eps_\textnormal{QSC})-3\eps_{\textnormal{QSC}}\log_2{3}}{2}
\end{IEEEeqnarray*} 
is the capacity (in $4$-ary units) of $\QSC{\eps_\textnormal{QSC}}$~\cite{WeidmannLechner12_1}. We are aiming to have $\const{M}'>\const{M}$ for a sufficiently large blocklength, and this is equivalent to
\begin{IEEEeqnarray*}{rCl}
    \IEEEeqnarraymulticol{3}{l}{%
    \const{M}'=4^{n'\const{C}_\textnormal{QSC}}\stackrel{(a)}{=}2^{2\cdot\frac{n}{3}\const{C}_\textnormal{QSC}}
    >\const{M}=2^{n\const{C}_\textnormal{BSC}}}\\*\hspace*{6mm}%
    & \iff &\frac{2}{3}\const{C}_\textnormal{QSC}>\const{C}_\textnormal{BSC},\IEEEeqnarraynumspace
\end{IEEEeqnarray*}
where $(a)$ holds since $n=3$ channel uses are required to construct an equivalent QSC via the ADC. However, Fig.~\ref{fig:capacities_ADCequivalent_qSCs} demonstrates that it is only possible to have a slightly larger $\frac{2}{3}\const{C}_\textnormal{QSC}>\const{C}_\textnormal{BSC}$ when $\gamma$ is larger than approximately $0.93$.

\ifthenelse{\boolean{JOURNAL}}{
\section{Performance Analysis for Classical Communication Over the EA ADC}
\label{sec:performance-analysis_EA-CQADC}

\todo[inline,author=Lin]{The following text lacks organization and contains some redundancy.}
\subsection{Good Code With Four Codewords for Single-Use EA ADC}
\label{sec:good-code_EA-cqADCn1}

\begin{theorem}
  \label{thm:best-pure-rho_EA-cqADC_single-use}
\end{theorem}

\section{New Finite-Blocklength Performance Bounds\\ Over the $q$-ary Asymmetric Channel ($q$-AC)}
\label{sec:new-FBL-bounds_qAC}

Inspired by Sections~\ref{sec:motivating-example} and \ref{sec:asymptotic-analysis}, we further proceed with making use the benefits of a collective measurement for the nonasymptotic performance analysis. Since performing a collective measurement using the $[3,2]_2$ SPC code and an associated $\code{S}^{\star}_{\textnormal{pm}}=\{\rho_0=\ketbra{+}{+}, \rho_1=\ketbra{-}{-}\}$ over the ADC induces a QSC, we are wondering if we perform such a collective measurement multiple times can lead to any benefits at finite blocklengths.
}

\ifthenelse{\boolean{NOTE}}{
{\hy 
\begin{theorem}[Upper Bound for the induced $q$-AC]
  \label{thm:UB_qAC}
  Consider the classical $q$-AC that induced from an EA ADC with arbitrary damming probability $0\leq\gamma\leq 1$. For any $(n, \const{M})_q$ code $\code{C}$ with the associated Bell input states, the average success probability using $\set{D}^{\textnormal{ML}}$ defined in Theorem~\ref{thm:uncoding_not-help_ADC} is bounded from above by
  \begin{IEEEeqnarray}{rCl}
    \label{eq:UB_qAC}
    P_{\textnormal{c}}(\code{C},\set{D}^{\textnormal{ML}})& \leq &\const{A}_{\vect{s}^\ast}\prod_{i=1}^{q-1} p_i^{s^\ast_{i-1} - s^\ast_i}+\sum_{i=1}^{q-1}\Biggl[\prod_{l=1}^{i-1}\binom{s^\ast_{l-1}}{s^\ast_l} p_l^{s^\ast_{l-1} - s^\ast_{l}}\Biggr]
    \nonumber\\
    &&\hspace*{0.25cm}\>\cdot\Biggl[\sum_{j=0}^{s^\ast_{i}-1}\binom{s^\ast_{i-1}}{j} p_i^{s^\ast_{i-1}-j}\biggl(1 - \sum_{l=1}^i p_{l}\biggr)^j\Biggr],
    \IEEEeqnarraynumspace
  \end{IEEEeqnarray}
  where for $\vect{s}\in\set{T}_{n,q}\eqdef\bigl\{\vect{s}=(s_0,s_1,\ldots,s_{q})\in [0:n]^{q+1}: s_0 = n,\,s_i\geq s_{i+1},\,i\in [0:q-1]\bigr\}$,
  \begin{IEEEeqnarray}{rCl}
    \const{A}_{\vect{s}}& \eqdef &\frac{q^n}{\const{M}}-\sum_{i=1}^{q-1} 
    \Biggl[\prod_{l=1}^{i-1} \binom{s_{l-1}}{s_l}\Biggr]
    \Biggl[\sum_{j=0}^{s_{i}-1} \binom{s_{i-1}}{j} (q-i)^j\Biggr],\IEEEeqnarraynumspace
  \end{IEEEeqnarray}
  and $\vect{s}^\ast\eqdef\argmin_{\vect{s}\in\set{T}_{n,q}}\{\const{A}_{\vect{s}}\colon\const{A}_{\vect{s}} \geq 0\}$.
  \IEEEeqnarraynumspace
\end{theorem}
\begin{IEEEproof}[Proof 1] (Sphere packing)
  For any message, the decoding region contains $\frac{q^n}{\MMM}$ erroneous codewords, because of the symmetries of the channel and initial distribution. We fill up this region with the most likely errors to get the highest success probability, regardless of whether it is possible in the given space. Because of the assumption on the transition probabilities $\vect{a}$ having more flips in the received codeword is less likely independently of the kind of the flips. Thus a greedy selection of errors gives optimality.
\end{IEEEproof}

\begin{IEEEproof}[Proof 2] (Hypothesis testing)
  We create a Neyman-Pearson test for fixed type-II error $\beta$ and with optimal type-I error $\alpha$. Similarly to \cite[Th.~35]{PolyanskiyPoorVerdu10_1}, $H_0$ is distributed as the asymmetric errors of a codeword with length $n$, and $H_1$ is uniform on these errors. To arrive at a minimal $\alpha$, we need to maximize $\beta$, which means $\beta = \nicefrac{1}{\MMM}$ (see \cite[Th.~28]{PolyanskiyPoorVerdu10_1}). By recursively choosing the errors with the highest probability wrt $H_0$, $A_{\vect{t}}$ and $\vect{t}$ is given via the equation for the type-II error:
  \begin{IEEEeqnarray}{rCl} \label{eq:beta_At}
    \frac{1}{\MMM}& = &\sum_{i=1}^{q-1} 
    \prod_{l=1}^{i-1} \binom{t_{l-1}}{t_l} 
    \sum_{j=0}^{t_i-1} \binom{t_{i-1}}{j} \frac{(q-i)^j}{q^n} + \frac{A_{\vect{t}}}{q^n}.
    \IEEEeqnarraynumspace
  \end{IEEEeqnarray}
  Then $P_{\textnormal{c}}(\code{C}) \leq 1 - \alpha$, which can be calculated by summing over the same errors as in \ref{eq:beta_At} but wrt to $H_0$ instead of $H_1$.
\end{IEEEproof}

\begin{remark}
  The assumption $a_j \in [a_i^2, a_i]$ if $j \geq i$ simplifies these bounds and was always fulfilled in our ADC simulations.
\end{remark}

\begin{theorem}[Random Coding Bound]
  \label{thm:LB_qAC}
  Let $p_e$ be the probability of an error $e$ and $n_e$ the number of errors with the same probability. Furthermore, let $\code{C}^{(n,\MMM)_q)}$ be a code, where all the codewords are chosen uniformly with repetition. Then the expected average success probability is
  \begin{IEEEeqnarray*}{rCl}
    \expect{P_{\textnormal{c}}(\code{C}^{(n,\MMM)_q)}, \set{D}^{\textnormal{ML}})}& = & 
    \sum_{e} p_e \sum_{0 \leq l < \MMM} \frac{q^{-n(\MMM-1)}}{l+1} 
    \IEEElinebreak{\cdot} \binom{\MMM-1}{l} n_e^l \paren{\sum_{p_{e'} < p_e} n_{e'}}^{\MMM-1-l}.
  \end{IEEEeqnarray*}
\end{theorem}
\begin{IEEEproof}
  This is a special case of \cite[Th.~15]{PolyanskiyPoorVerdu10_1}.
\end{IEEEproof}

\subsection{Some Results for $q$-SC}
\label{sec:results_qSC}

\begin{theorem}[RCB for $q$-SC]
  \label{thm:rcb_q-sc}
  Consider a $q$-SC with cross-over parameter $\eps \in [0, 1-\nicefrac{1}{q}]$, the success probability of a random $(\const{M},n)_q$ code is 
  \begin{IEEEeqnarray}{rCl}
    P_{\textnormal{c}}(\code{C}) &=& \sum_{i=0}^{n} \binom{n}{i} \eps^i (1-\eps)^{n-i}
    \IEEElinebreak{\cdot} \sum_{l=0}^{\const{M}-1} \binom{\const{M}-1}{l} \frac{p_i^l \cdot s_i^{\const{M} - 1 - l}}{l+1},
  \end{IEEEeqnarray}
  where $p_i\eqdef\frac{\binom{n}{i} (q-1)^i}{q^n}$ and $s_i\eqdef\sum_{j > i} p_j$.
\end{theorem}

\begin{theorem}[PPV Converse]  
  \label{thm:converse_bound_q-arySC}
  Consider a $q$-SC with crossover probability $0\leq\veps\leq\nicefrac{1}{q}$. Then, the maximum possible average success probability of any $(\const{M},n)_q$ code $\code{C}$ is bounded from above by
  \begin{IEEEeqnarray}{rCl} \label{eq:converse_bound_q-arySC}
    P_{\text{c}}(\code{C}) &\leq& \sum_{j=0}^{t-1} \binom{n}{j} (q-1)^j \cdot \veps^{j} (1-(q-1)\veps)^{n-j}
    \nonumber\\
    && +\> \const{A}_t \veps^t (1-(q-1)\veps)^{n-t},
  \end{IEEEeqnarray}
  where $t$ is the largest integer such that
  \begin{IEEEeqnarray*}{c}
    \const{A}_t\eqdef\frac{q^n}{\const{M}} - \sum_{j=0}^{t-1} \binom{n}{j}(q-1)^j \geq 0.
  \end{IEEEeqnarray*}
\end{theorem}
}}{}

\ifthenelse{\boolean{JOURNAL}}{
  \input{section_main_results.tex}
}{}

\section{Numerical Results}
\label{sec:numerical-results}

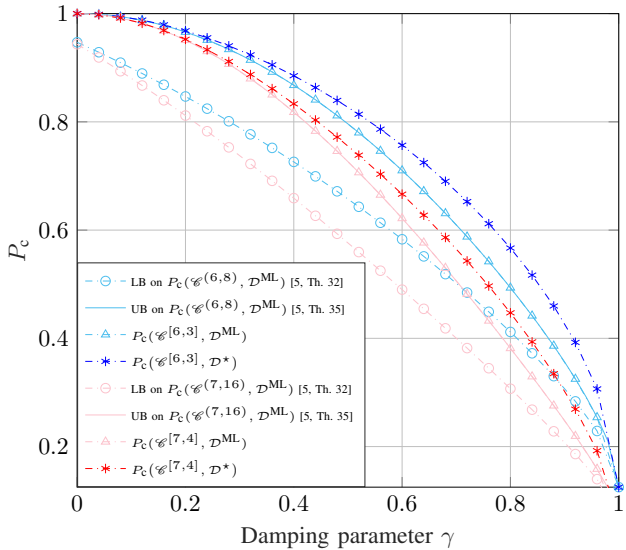
\begin{figure}[t!]
  \centering
  \input{\Figs/ADC-success_prob_vs_gamma_n6_k3_q2.tex}
  \caption{Performance bounds on the average success probability of the $[n,k]_2=[6,3]_2$ reduced Hamming code and the $[7,4]_2$ Hamming code over the ADC $\set{N}_\gamma$ for  $0\leq\gamma\leq 1$.} 
  \label{fig:performance_bounds_ADCequivalent_n6_k3_q2}
  \vspace{-2ex}
\end{figure}

In this section, we give another example to demonstrate the benefit of using collective measurements on the outputs over the ADC, illustrated in Fig.~\ref{fig:performance_bounds_ADCequivalent_n6_k3_q2}. We select two binary codes, the $[6,3]_2$ reduced Hamming code and the $[7,4]_2$ Hamming code with generator matrices
\begin{equation*}
  \Scale[1.0]{\begin{bmatrix}
    1 & 0 & 0 & 1 & 1 & 0 \\[-.75mm]
    0 & 1 & 0 & 1 & 0 & 1 \\[-.75mm]
    0 & 0 & 1 & 0 & 1 & 1
  \end{bmatrix}},\quad
\Scale[1.0]{\begin{bmatrix}
  1 & 0 & 0 & 0 & 0 & 1 & 1 \\[-.75mm]
    0 & 1 & 0 & 0 & 1 & 0 & 1 \\[-.75mm]
    0 & 0 & 1 & 0 & 1 & 1 & 0 \\[-.75mm]
    0 & 0 & 0 & 1 & 1 & 1 & 1
  \end{bmatrix}},
\end{equation*}
respectively, and use the associated $\code{S}^\star_{\textnormal{pm}}$ for encoding. One can observe a similar quantum performance gain as in Fig.~\ref{fig:performance_bounds_ADCequivalent_qSC}.

\ifthenelse{\boolean{JOURNAL}}{
\begin{figure}[t!]
  \centering
  \input{\Figs/ADC-success_prob_vs_gamma_n6_k4_q2}
  \caption{Performance bounds on the success probability of $[n,k]_q=[6,4]_2$ code reduced from the $[6,5]_2$ cyclic code for the $\ADC{\gamma}$.}
  \label{fig:performance_bounds_ADCequivalent_qSC}
\end{figure}
  
\begin{figure}[t!]
  \centering
  \input{\Figs/EA_ADC-success_prob_vs_gamma_n3_k1_q4}
  \caption{Performance bounds on the success probability of the $[n,k]_q=[3,1]_4$ repetition code for $\set{N}_{\gamma} \otimes \III$.}
  \label{fig:performance_bounds_ADCequivalent_qSC}
\end{figure}
}{}

\section{Conclusion and Future Work}
\label{sec:conclusion}

In this work, we studied practical finite-blocklength CQ channel coding over a noisy CQ channel modeled by amplitude damping errors, without shared entanglement between the sender and receiver.
 We proved that using an uncoded approach with an associated set of input states that achieves the optimal average success probability for a single-use ADC, there will be no quantum performance gain even if a collective-measurement strategy is performed for any blocklength. However, once the coded transmission of classical data is employed, the collective-measurement strategy can outperform the strategy of performing individual measurements. We provided numerical results to support these findings.  

Moving forward, a natural next step is to address the finite-blocklength regime with large code sizes over the quantum ADC using the collective-measurement strategy. However, since fully performing collective measurements on all the channel outputs is computationally intensive, we plan to design a hybrid approach that first applies collective measurements on partial channel outputs, and then perform individual measurements over the resulting induced channel. This, as a result, leads us to consider a multiple input and potentially nonsymmetric induced DM channel. As a concrete example, we may analyze the finite-blocklength coding performance over the QSC in Section~\ref{sec:motivating-example}. Our goal is to investigate whether such a hybrid coding strategy offers practical advantages within the finite-blocklength regime.

\section*{Acknowledgment}

The authors would like to thank Hao-Chung Cheng for his valuable discussions on various system setups for classical communication over quantum channels.


\ifthenelse{\boolean{arXiv}}{\IEEEtriggeratref{14}}{
  \IEEEtriggeratref{14} 
} 
\appendices

\ifthenelse{\boolean{arXiv}}{\input{proof_Theorem1.tex}}{}

\ifthenelse{\boolean{JOURNAL}}{\input{proof_lem_valid_EA_inputs}}{}

\ifthenelse{\boolean{arXiv}}{{\input{proof_thm_n=k_ADC}}}{}

\ifthenelse{\boolean{JOURNAL}}{\input{proof_thm_n=k_combined}}{}

\bibliographystyle{IEEEtran}
\bibliography{defshort1, biblioHY}

\end{document}

%% file: figs/coding_cqadc_M2n3_indep.tex
\begin{tikzpicture}[scale=0.9, y=-1cm]
  \foreach\i in {1,2,3}{
    \draw[line width=0.3mm] (-1.8,2) node[left]{$m$}--(-1.3,2);
    \draw[line width=0.3mm] (-1.3,1)--(-1.3,3);
    \draw[line width=0.3mm] (-1.3, \i) node[right,yshift=2.15mm,xshift=3.0mm] {$x_{m,\i}$}-- (0.6, \i);
    \draw[line width=0.3mm] (0.6, \i) node[draw,fill=white] {$\rho_{x_{m,\i}}$};
    \draw[blue, line width=0.3mm] (1.17, \i) -- (2.1,\i);
    \draw[line width=0.3mm] (2.1, \i) node[draw,fill=white] {$\set{N}$};
    \draw[magenta, line width=0.3mm] (2.42, \i) -- (3.1, \i);
    \draw[line width=0.3mm] (3.7, \i) node[right,yshift=2.15mm,xshift=1mm] {$y_\i$}-- (4.7, \i);
    \draw[line width=0.3mm] (4.70,1)--(4.70,3);
    \draw[line width=0.3mm] (4.70,2) -- (5.5, 2) node[draw,fill=white, rectangle, minimum width=1cm, minimum height=1cm] {$g$}--(6.3,2) node[right,yshift=.5mm]{$\hat{m}$};
  }
  \draw[line width=0.3mm] (3.4, 1) node[draw,fill=white] {$\set{D}$};
  \draw[line width=0.3mm] (3.4, 2) node[draw,fill=white] {$\set{D}$};
  \draw[line width=0.3mm] (3.4, 3) node[draw,fill=white] {$\set{D}$};
\end{tikzpicture}

%% file: figs/coding_cqadc_M2n3_joint_outputs.tex
\begin{tikzpicture}[scale=0.9, y=-1cm]
  \foreach\i in {1,2,3}{
    \draw[line width=0.3mm] (-1.8,2) node[left]{$m$}--(-1.3,2);
    \draw[line width=0.3mm] (-1.3,1)--(-1.3,3);
    \draw[line width=0.3mm] (-1.3, \i) node[right,yshift=2.15mm,xshift=3.0mm] {$x_{m,\i}$}-- (0.6, \i);
    \draw[line width=0.3mm] (0.6, \i) node[draw,fill=white] {$\rho_{x_{m,\i}}$};
    \draw[blue, line width=0.3mm] (1.17, \i) -- (2.1,\i);
    \draw[line width=0.3mm] (2.1, \i) node[draw,fill=white] {$\set{N}$};
    \draw[magenta, line width=0.3mm] (2.42, \i) -- (4.4, \i);
    \draw[line width=0.3mm] (4.4, 2) node[draw,fill=white, minimum height=2.5cm, minimum width=2cm] {{$\set{D}^\star$}}-- (6.3,2) node[right,yshift=.5mm]{$\hat{m}$};
  }
\end{tikzpicture}

%% file: figs/success_prob_versus_gamma_cqadc_M4n3_v1.tex
%
%
\definecolor{mycolor1}{rgb}{0.30100,0.74500,0.93300}%
\begin{tikzpicture}

\begin{axis}[%
width=7.75cm,
height=6.50cm,
at={(1.011in,0.642in)},
scale only axis,
xmin=0,
xmax=1,
xlabel style={font=\color{white!15!black}},
xlabel={Damping parameter $\gamma$},
ymin=0.25,
ymax=1.00,
yminorticks=true,
ylabel style={font=\color{white!15!black}},
ylabel={$P_{\textnormal{c}}$},
ylabel style = {yshift=-1mm},
axis background/.style={fill=white},
xmajorgrids,
ymajorgrids,
legend style={at={(0.00,0.00)}, anchor=south west, legend cell align=left, align=left, draw=white!15!black, font=\scriptsize}
]

\addplot [color=mycolor1, dashdotted, mark size=2.0pt, mark=o, mark options={solid, mycolor1}, each nth point=2, filter discard warning=false]
table[row sep=crcr]{%
0	0.827636718749938\\
0.01	0.824042056226208\\
0.02	0.820435691593154\\
0.03	0.816817454521701\\
0.04	0.8131871703291\\
0.05	0.80954465982053\\
0.06	0.805889739123194\\
0.07	0.802222219512458\\
0.08	0.7985419072296\\
0.09	0.794848603290621\\
0.1	0.791142103285611\\
0.11	0.787422197168075\\
0.12	0.783688669033598\\
0.13	0.7799412968872\\
0.14	0.776179852398637\\
0.15	0.772404100644894\\
0.16	0.768613799839042\\
0.17	0.764808701044553\\
0.18	0.760988547874111\\
0.19	0.757153076171893\\
0.2	0.753302013678165\\
0.21	0.749435079675004\\
0.22	0.74555198461181\\
0.23	0.741652429709216\\
0.24	0.737736106539773\\
0.25	0.733802696583812\\
0.26	0.729851870758616\\
0.27	0.72588328891891\\
0.28	0.721896599326516\\
0.29	0.717891438086818\\
0.3	0.71386742854944\\
0.31	0.709824180670332\\
0.32	0.705761290332178\\
0.33	0.701678338619749\\
0.34	0.697574891046476\\
0.35	0.693450496728197\\
0.36	0.689304687499561\\
0.37	0.685136976968175\\
0.38	0.680946859501027\\
0.39	0.676733809137159\\
0.4	0.672497278419925\\
0.41	0.66823669714144\\
0.42	0.663951470990995\\
0.43	0.659640980098314\\
0.44	0.655304577461455\\
0.45	0.650941587247998\\
0.46	0.646551302956794\\
0.47	0.64213298542601\\
0.48	0.637685860671477\\
0.49	0.633209117537282\\
0.5	0.628701905138314\\
0.51	0.624163330071739\\
0.52	0.619592453371374\\
0.53	0.614988287175354\\
0.54	0.610349791073378\\
0.55	0.605675868095042\\
0.56	0.600965360295155\\
0.57	0.596217043885406\\
0.58	0.591429623854038\\
0.59	0.5866017280061\\
0.6	0.581731900346105\\
0.61	0.57681859371213\\
0.62	0.571860161555137\\
0.63	0.566854848739005\\
0.64	0.56180078121473\\
0.65	0.556695954395613\\
0.66	0.551538220027849\\
0.67	0.54632527131132\\
0.68	0.541054625976723\\
0.69	0.535723606964971\\
0.7	0.530329320279927\\
0.71	0.524868629491781\\
0.72	0.519338126250094\\
0.73	0.513734096015348\\
0.74	0.508052478025387\\
0.75	0.502288818264667\\
0.76	0.496438213880316\\
0.77	0.490495247062673\\
0.78	0.484453905840904\\
0.79	0.478307488481309\\
0.8	0.472048487136704\\
0.81	0.46566844496053\\
0.82	0.459157778889409\\
0.83	0.45250555743645\\
0.84	0.445699218691192\\
0.85	0.438724207594662\\
0.86	0.431563502303186\\
0.87	0.424196985129771\\
0.88	0.416600590758547\\
0.89	0.408745127000286\\
0.9	0.400594599637171\\
0.91	0.392103759765505\\
0.92	0.383214380959529\\
0.93	0.373849355668393\\
0.94	0.363902810591464\\
0.95	0.353222365994777\\
0.96	0.341574218390677\\
0.97	0.328564902837587\\
0.98	0.313427448115925\\
0.99	0.294182371734642\\
1	0.249999999992301\\
};
\addlegendentry{{LB on $P_{\textnormal{c}}(\code{C}^{(3,4)},\set{D}^{\textnormal{ML}})$~\cite[Th.~32]{PolyanskiyPoorVerdu10_1}}}

\addplot [color=mycolor1]
table[row sep=crcr]{%
0	0.999999999999913\\
0.01	0.994993718553225\\
0.02	0.9899747468305\\
0.03	0.984942890089724\\
0.04	0.979897948556558\\
0.05	0.974839717240372\\
0.06	0.969767985741561\\
0.07	0.96468253804958\\
0.08	0.959583152331209\\
0.09	0.954469600708416\\
0.1	0.949341649025205\\
0.11	0.944199056602785\\
0.12	0.939041575982305\\
0.13	0.93386895265441\\
0.14	0.928680924774763\\
0.15	0.923477222864631\\
0.16	0.91825756949558\\
0.17	0.913021678957221\\
0.18	0.907769256906886\\
0.19	0.902500000000025\\
0.2	0.897213595499993\\
0.21	0.891909720865825\\
0.22	0.886588043316438\\
0.23	0.881248219369651\\
0.24	0.875889894354107\\
0.25	0.870512701892251\\
0.26	0.865116263352151\\
0.27	0.859700187265879\\
0.28	0.854264068711908\\
0.29	0.848807488658767\\
0.3	0.843330013266948\\
0.31	0.837831193145766\\
0.32	0.832310562561568\\
0.33	0.826767638593351\\
0.34	0.821201920231438\\
0.35	0.815612887414461\\
0.36	0.809999999999406\\
0.37	0.804362696658943\\
0.38	0.798700393699666\\
0.39	0.793012483794198\\
0.4	0.787298334619361\\
0.41	0.781557287391762\\
0.42	0.775788655291193\\
0.43	0.769991721761146\\
0.44	0.764165738674555\\
0.45	0.758309924351426\\
0.46	0.752423461413523\\
0.47	0.746505494459387\\
0.48	0.740555127540975\\
0.49	0.734571421420821\\
0.5	0.728553390585928\\
0.51	0.722499999991487\\
0.52	0.716410161503937\\
0.53	0.71028273000871\\
0.54	0.70411649914322\\
0.55	0.697910196610004\\
0.56	0.691662479018412\\
0.57	0.68537192619554\\
0.58	0.679037034898107\\
0.59	0.67265621184631\\
0.6	0.666227765988111\\
0.61	0.659749899887424\\
0.62	0.653220700111783\\
0.63	0.646638126473652\\
0.64	0.639999999953705\\
0.65	0.633303989103195\\
0.66	0.626547594684528\\
0.67	0.619728132262763\\
0.68	0.612842712403654\\
0.69	0.605888218063332\\
0.7	0.598861278666905\\
0.71	0.591758240263343\\
0.72	0.584575131005329\\
0.73	0.57730762102661\\
0.74	0.569950975563742\\
0.75	0.562499999877666\\
0.76	0.554948974150679\\
0.77	0.547291576034281\\
0.78	0.539520787858204\\
0.79	0.531628784615858\\
0.8	0.523606797622247\\
0.81	0.515444947057099\\
0.82	0.507132034247644\\
0.83	0.498655281187858\\
0.84	0.489999999925294\\
0.85	0.481149167255672\\
0.86	0.472082869303646\\
0.87	0.462777563754915\\
0.88	0.453205080750189\\
0.89	0.443331239516613\\
0.9	0.433113883008415\\
0.91	0.42249999999985\\
0.92	0.41142135623727\\
0.93	0.399787565552889\\
0.94	0.387474487139131\\
0.95	0.374303398661923\\
0.96	0.359999999559767\\
0.97	0.344102539703642\\
0.98	0.325710677080892\\
0.99	0.302499998424142\\
1	0.249999999990953\\
};
\addlegendentry{{UB on $P_{\textnormal{c}}(\code{C}^{(3,4)},\set{D}^{\textnormal{ML}})$~\cite[Th.~35]{PolyanskiyPoorVerdu10_1}}}

\addplot [color=mycolor1, dashdotted, mark=triangle, mark options={solid, mycolor1}, each nth point=2, filter discard warning=false]
  table[row sep=crcr]{%
0	0.999999999999913\\
0.01	0.994993718553225\\
0.02	0.9899747468305\\
0.03	0.984942890089724\\
0.04	0.979897948556558\\
0.05	0.974839717240373\\
0.06	0.969767985741561\\
0.07	0.96468253804958\\
0.08	0.959583152331209\\
0.09	0.954469600708415\\
0.1	0.949341649025205\\
0.11	0.944199056602785\\
0.12	0.939041575982305\\
0.13	0.93386895265441\\
0.14	0.928680924774763\\
0.15	0.923477222864631\\
0.16	0.91825756949558\\
0.17	0.91302167895722\\
0.18	0.907769256906885\\
0.19	0.902500000000025\\
0.2	0.897213595499993\\
0.21	0.891909720865825\\
0.22	0.886588043316438\\
0.23	0.881248219369651\\
0.24	0.875889894354108\\
0.25	0.870512701892251\\
0.26	0.865116263352151\\
0.27	0.859700187265879\\
0.28	0.854264068711908\\
0.29	0.848807488658767\\
0.3	0.843330013266949\\
0.31	0.837831193145766\\
0.32	0.832310562561568\\
0.33	0.826767638593351\\
0.34	0.821201920231438\\
0.35	0.815612887414461\\
0.36	0.809999999999406\\
0.37	0.804362696658943\\
0.38	0.798700393699666\\
0.39	0.793012483794198\\
0.4	0.787298334619361\\
0.41	0.781557287391762\\
0.42	0.775788655291193\\
0.43	0.769991721761146\\
0.44	0.764165738674555\\
0.45	0.758309924351426\\
0.46	0.752423461413523\\
0.47	0.746505494459387\\
0.48	0.740555127540975\\
0.49	0.734571421420821\\
0.5	0.728553390585928\\
0.51	0.722499999991487\\
0.52	0.716410161503937\\
0.53	0.71028273000871\\
0.54	0.70411649914322\\
0.55	0.697910196610004\\
0.56	0.691662479018411\\
0.57	0.68537192619554\\
0.58	0.679037034898107\\
0.59	0.67265621184631\\
0.6	0.666227765988111\\
0.61	0.659749899887424\\
0.62	0.653220700111783\\
0.63	0.646638126473653\\
0.64	0.639999999953705\\
0.65	0.633303989103194\\
0.66	0.626547594684528\\
0.67	0.619728132262763\\
0.68	0.612842712403654\\
0.69	0.605888218063332\\
0.7	0.598861278666905\\
0.71	0.591758240263343\\
0.72	0.584575131005329\\
0.73	0.57730762102661\\
0.74	0.569950975563742\\
0.75	0.562499999877666\\
0.76	0.554948974150679\\
0.77	0.547291576034281\\
0.78	0.539520787858204\\
0.79	0.531628784615858\\
0.8	0.523606797622247\\
0.81	0.515444947057099\\
0.82	0.507132034247644\\
0.83	0.498655281187859\\
0.84	0.489999999925295\\
0.85	0.481149167255672\\
0.86	0.472082869303646\\
0.87	0.462777563754914\\
0.88	0.453205080750189\\
0.89	0.443331239516613\\
0.9	0.433113883008415\\
0.91	0.42249999999985\\
0.92	0.41142135623727\\
0.93	0.39978756555289\\
0.94	0.387474487139131\\
0.95	0.374303398661923\\
0.96	0.359999999559767\\
0.97	0.344102539703642\\
0.98	0.325710677080892\\
0.99	0.302499998424142\\
1	0.250000000009047\\
};
\addlegendentry{$P_{\textnormal{c}}(\code{C}_\textnormal{SPC}^{[3,2]},\set{D}^{\textnormal{ML}})$}

\addplot [color=blue, dashdotted, mark=asterisk, mark options={solid, blue}, each nth point=2, filter discard warning=false]
  table[row sep=crcr]{%
0	0.999999998502896\\
0.01	0.995043340097692\\
0.02	0.990171692464258\\
0.03	0.985382491959733\\
0.04	0.980673091698525\\
0.05	0.976040766281638\\
0.06	0.971482715367352\\
0.07	0.966996067557616\\
0.08	0.962577884215457\\
0.09	0.958225164588336\\
0.1	0.953934849652402\\
0.11	0.949703826865223\\
0.12	0.945528934556103\\
0.13	0.941406965387089\\
0.14	0.937334663972224\\
0.15	0.933308887265619\\
0.16	0.929326050816047\\
0.17	0.925383133926698\\
0.18	0.92147662936772\\
0.19	0.917603188851614\\
0.2	0.913759440344343\\
0.21	0.909941942889502\\
0.22	0.906147503405322\\
0.23	0.902372552639577\\
0.24	0.89861378051496\\
0.25	0.89486781851781\\
0.26	0.891131308628115\\
0.27	0.887400906091659\\
0.28	0.883673281436906\\
0.29	0.879945122362017\\
0.3	0.876213135579203\\
0.31	0.872474047717854\\
0.32	0.868724606565943\\
0.33	0.864961581415178\\
0.34	0.861181763521958\\
0.35	0.857381965794463\\
0.36	0.853559022534902\\
0.37	0.849709788510487\\
0.38	0.845831137887778\\
0.39	0.841919962788088\\
0.4	0.837973171292688\\
0.41	0.833987685389757\\
0.42	0.829960438398981\\
0.43	0.825888372214561\\
0.44	0.821768434710636\\
0.45	0.817597574165038\\
0.46	0.813372717048534\\
0.47	0.809090848519005\\
0.48	0.804748814578434\\
0.49	0.800343612790869\\
0.5	0.795872078689505\\
0.51	0.791331059616343\\
0.52	0.786717377199394\\
0.53	0.782027799942558\\
0.54	0.777259041160602\\
0.55	0.772407750465871\\
0.56	0.767470503407678\\
0.57	0.762443790297232\\
0.58	0.757324004009366\\
0.59	0.752107426443682\\
0.6	0.746790213765729\\
0.61	0.741368380075009\\
0.62	0.735837779461543\\
0.63	0.730194085721384\\
0.64	0.724432770166157\\
0.65	0.718549075330791\\
0.66	0.712537991403611\\
0.67	0.706394221631968\\
0.68	0.700112147088208\\
0.69	0.69368578715331\\
0.7	0.687108754063721\\
0.71	0.680374200802045\\
0.72	0.673474761360777\\
0.73	0.6664024817051\\
0.74	0.659148739634522\\
0.75	0.651704151204833\\
0.76	0.644058460820765\\
0.77	0.636200411075432\\
0.78	0.628117587720181\\
0.79	0.619796233633015\\
0.8	0.61122102331719\\
0.81	0.602374784014424\\
0.82	0.593238263122462\\
0.83	0.583789331984565\\
0.84	0.574002971235045\\
0.85	0.563850253747025\\
0.86	0.553297532604149\\
0.87	0.542305264397721\\
0.88	0.530826420444453\\
0.89	0.518804295915416\\
0.9	0.506169436282819\\
0.91	0.492835033130239\\
0.92	0.478690245592185\\
0.93	0.463589299399261\\
0.94	0.447333561184561\\
0.95	0.42963940586275\\
0.96	0.410075223386937\\
0.97	0.387920874530307\\
0.98	0.361785139215467\\
0.99	0.328111273422971\\
1	0.249999998071716\\
};
\addlegendentry{$P_{\textnormal{c}}\bigl(\code{C}^{[3,2]}_{\textnormal{SPC}}, \set{D}^{\star}\bigr)$} 

\end{axis}
\end{tikzpicture}%

%% file: figs/CQADC_capacities.tex
%
%
%
\begin{tikzpicture}[spy using outlines={circle, magnification=50.0, size=3.0cm, connect spies}]

\begin{axis}[%
width=7.75cm,
height=6.50cm,
at={(1.011in,0.642in)},
scale only axis,
xmin=0,
xmax=1,
xlabel style={font=\color{white!15!black}},
xlabel={Damping parameter $\gamma$},
ymin=0,
ymax=1,
axis background/.style={fill=white},
xmajorgrids,
ymajorgrids,
legend style={at={(0.00,0.00)}, anchor=south west, legend cell align=left, align=left, draw=white!15!black}
]
\addplot [very thin, color=blue]
  table[row sep=crcr]{%
0	0.999999999998014\\
0.01	0.974733872272511\\
0.02	0.954392551002323\\
0.03	0.935876564901779\\
0.04	0.918531084985385\\
0.05	0.9020465015803\\
0.06	0.886241213985096\\
0.07	0.870995699559106\\
0.08	0.856225330713904\\
0.09	0.841867063439635\\
0.1	0.827872137216044\\
0.11	0.814201733717834\\
0.12	0.800824232221002\\
0.13	0.787713389199682\\
0.14	0.774847083283223\\
0.15	0.762206421942239\\
0.16	0.74977508838892\\
0.17	0.737538853080353\\
0.18	0.725485201080166\\
0.19	0.7136030428841\\
0.2	0.701882486605515\\
0.21	0.690314656085391\\
0.22	0.678891543925392\\
0.23	0.667605891458788\\
0.24	0.656451089767579\\
0.25	0.645421097334795\\
0.26	0.634510370984616\\
0.27	0.623713807539324\\
0.28	0.61302669419622\\
0.29	0.602444666057624\\
0.3	0.591963669572874\\
0.31	0.581579930900623\\
0.32	0.57128992839271\\
0.33	0.561090368551315\\
0.34	0.550978164929603\\
0.35	0.540950419540046\\
0.36	0.531004406409673\\
0.37	0.521137556981961\\
0.38	0.511347447114034\\
0.39	0.501631785457704\\
0.4	0.491988403045636\\
0.41	0.482415243930845\\
0.42	0.472910356750087\\
0.43	0.463471887100294\\
0.44	0.454098070632751\\
0.45	0.444787226782809\\
0.46	0.435537753063929\\
0.47	0.426348119864207\\
0.48	0.417216865691466\\
0.49	0.408142592819793\\
0.5	0.3991239632962\\
0.51	0.390159695271068\\
0.52	0.381248559620351\\
0.53	0.372389376831231\\
0.54	0.363581014126133\\
0.55	0.354822382802832\\
0.56	0.346112435770806\\
0.57	0.337450165266161\\
0.58	0.328834600729291\\
0.59	0.320264806831125\\
0.6	0.311739881635224\\
0.61	0.303258954884307\\
0.62	0.294821186400889\\
0.63	0.286425764592742\\
0.64	0.278071905054769\\
0.65	0.269758849259677\\
0.66	0.26148586333054\\
0.67	0.253252236888998\\
0.68	0.245057281973355\\
0.69	0.236900332021417\\
0.7	0.228780740913288\\
0.71	0.220697882069819\\
0.72	0.212651147602729\\
0.73	0.204639947512785\\
0.74	0.196663708932683\\
0.75	0.188721875411604\\
0.76	0.1808139062386\\
0.77	0.172939275802248\\
0.78	0.165097472984171\\
0.79	0.157288000584219\\
0.8	0.14951037477529\\
0.81	0.141764124585899\\
0.82	0.134048791408767\\
0.83	0.126363928533809\\
0.84	0.118709100704079\\
0.85	0.111083883693301\\
0.86	0.103487863903796\\
0.87	0.0959206379837251\\
0.88	0.088381812462627\\
0.89	0.0808710034041713\\
0.9	0.0733878360745808\\
0.91	0.0659319446244058\\
0.92	0.05850297178344\\
0.93	0.051100568571055\\
0.94	0.0437243940221282\\
0.95	0.0363741148057752\\
0.96	0.0290494053307315\\
0.97	0.0217499471417831\\
0.98	0.0144754287627534\\
0.99	0.00722554559744504\\
1	0\\
};
\addlegendentry{$\const{C}_{\textnormal{BSC}}$}

\addplot [very thin, color=magenta]
  table[row sep=crcr]{%
0	0.666666650526469\\
0.01	0.6490201442627\\
0.02	0.63492324531726\\
0.03	0.622262476392223\\
0.04	0.610574369134477\\
0.05	0.599632649465373\\
0.06	0.589300149291879\\
0.07	0.57948345482594\\
0.08	0.570114051865252\\
0.09	0.561139020913606\\
0.1	0.552515892701141\\
0.11	0.544209567386633\\
0.12	0.536190352369285\\
0.13	0.528432650234408\\
0.14	0.520914037389564\\
0.15	0.513614894744781\\
0.16	0.506516920570065\\
0.17	0.499604571782621\\
0.18	0.492862899471474\\
0.19	0.486278479654355\\
0.2	0.479838909811478\\
0.21	0.473532594047704\\
0.22	0.467349146497322\\
0.23	0.461278271529606\\
0.24	0.45531072580745\\
0.25	0.449437730076579\\
0.26	0.443651025174809\\
0.27	0.437942827860565\\
0.28	0.432305792348303\\
0.29	0.426732977373723\\
0.3	0.421217817969404\\
0.31	0.415754099873817\\
0.32	0.410335937887265\\
0.33	0.404957755859974\\
0.34	0.399614269403122\\
0.35	0.394300469712053\\
0.36	0.389011609422339\\
0.37	0.383743189247947\\
0.38	0.378490945973971\\
0.39	0.373250841309415\\
0.4	0.368019051253902\\
0.41	0.362791956538371\\
0.42	0.357566133429212\\
0.43	0.35233834528116\\
0.44	0.347105535204297\\
0.45	0.341864815826406\\
0.46	0.336613439491587\\
0.47	0.33134889633162\\
0.48	0.326068669683795\\
0.49	0.320770592158515\\
0.5	0.315452464053516\\
0.51	0.310112267239125\\
0.52	0.304748121120082\\
0.53	0.299358253490813\\
0.54	0.293941002531628\\
0.55	0.28849481209978\\
0.56	0.283018225998107\\
0.57	0.277509882533147\\
0.58	0.271968509235958\\
0.59	0.266392917534064\\
0.6	0.260781997665691\\
0.61	0.255134713634524\\
0.62	0.249450098375179\\
0.63	0.243727248621404\\
0.64	0.237965320223014\\
0.65	0.232163522052841\\
0.66	0.226321115619224\\
0.67	0.220437407196715\\
0.68	0.214511743479284\\
0.69	0.208543507690096\\
0.7	0.202532115351733\\
0.71	0.196477010012077\\
0.72	0.190377659166283\\
0.73	0.184233550225922\\
0.74	0.178044186608992\\
0.75	0.171809083960776\\
0.76	0.165527766566471\\
0.77	0.159199763827451\\
0.78	0.152824606991831\\
0.79	0.146401826246462\\
0.8	0.139930947979525\\
0.81	0.133411490061273\\
0.82	0.126843036755712\\
0.83	0.120224971100723\\
0.84	0.113556850566079\\
0.85	0.106838168713511\\
0.86	0.100068416627334\\
0.87	0.0932470904428136\\
0.88	0.0863737013923336\\
0.89	0.079447790853052\\
0.9	0.0724689670470096\\
0.91	0.0654368924839674\\
0.92	0.0583514189497176\\
0.93	0.0512126057114376\\
0.94	0.0440208821175745\\
0.95	0.0367772728726293\\
0.96	0.0294837999369583\\
0.97	0.0221442795095782\\
0.98	0.014766119842125\\
0.99	0.00736551443569433\\
1	0\\
};
\addlegendentry{$\frac{2}{3}\const{C}_{\textnormal{QSC}}$}

\end{axis}
\spy [magenta] on (10.0,1.82) in node [left] at (9.5,5.5); 
\end{tikzpicture}%

%% file: figs/ADC-success_prob_vs_gamma_n6_k3_q2.tex



\definecolor{mycolor1}{rgb}{0.30100,0.74500,0.93300}
\definecolor{mycolor2}{rgb}{0.99, 0.76, 0.8}

\begin{tikzpicture}[scale=0.90]

\begin{axis}[
width=8.00cm,
height=7.00cm,
at={(1.011in,0.642in)},
scale only axis,
xmin=0, xmax=1, xlabel style={font=\color{white!15!black}}, xlabel={Damping parameter $\gamma$},
ymin=0.125, ymax=1.00, yminorticks=true, ylabel style={font=\color{white!15!black}}, 
ylabel={$P_{\textnormal{c}}$}, ylabel style = {yshift=-1mm}, 
axis background/.style={fill=white},
xmajorgrids, ymajorgrids, 
legend style={at={(0.00,0.00)}, anchor=south west, legend cell align=left, align=left, draw=white!15!black, font=\tiny} 
]
  
\addplot [color=mycolor1, dashdotted, mark size=2.0pt, mark=o, mark options={solid, mycolor1}, each nth point=2]
table[row sep=crcr]{
0.0 0.9469885022128873 \\
0.02 0.9379056678173546 \\
0.04 0.9286166094080908 \\
0.06 0.919121611410058 \\
0.08 0.9094207089300266 \\
0.1 0.8995140503637586 \\
0.12 0.8894016731779725 \\
0.14 0.8790835478981891 \\
0.16 0.8685595718191687 \\
0.18 0.8578295618911518 \\
0.2 0.8468932468610871 \\
0.22 0.8357502584437034 \\
0.24 0.8244001214213938 \\
0.26 0.8128422427164651 \\
0.28 0.8010758982922841 \\
0.3 0.7891002193766571 \\
0.32 0.7769141762171221 \\
0.34 0.7645165596905023 \\
0.36 0.7519059602918924 \\
0.38 0.7390807438824736 \\
0.4 0.7260390246099715 \\
0.42 0.7127786321158803 \\
0.44 0.6992970745015221 \\
0.46 0.6855914948198957 \\
0.48 0.6716586200257871 \\
0.5 0.6574947371415947 \\
0.52 0.6430954825310571 \\
0.54 0.6284559563411933 \\
0.56 0.613570492242156 \\
0.58 0.5984325635063392 \\
0.6 0.5830346309665609 \\
0.62 0.567367955548203 \\
0.64 0.5514223648792125 \\
0.66 0.5351859595629159 \\
0.68 0.5186447390458646 \\
0.7 0.5017821186652648 \\
0.72 0.4845782967950525 \\
0.74 0.4670094114153455 \\
0.76 0.44904639417263825 \\
0.78 0.4306533785282576 \\
0.8 0.411785431400981 \\
0.82 0.39238522077353327 \\
0.84 0.3723779417370288 \\
0.86 0.3516632187078432 \\
0.88 0.3301016837772682 \\
0.9 0.3074903081263216 \\
0.92 0.2835144971237549 \\
0.94 0.2576397013602689 \\
0.96 0.2288146643229345 \\
0.98 0.19430338536599445 \\
1.0 0.1249999999053821 \\
};
\addlegendentry{LB on $P_{\textnormal{c}}(\code{C}^{(6, 8)}, \set{D}^{\textnormal{ML}})$~\cite[Th.~32]{PolyanskiyPoorVerdu10_1}
}
  
\addplot [color=mycolor1]
table[row sep=crcr]{
0.0 0.9999999999999973 \\
0.02 0.9996509993495225 \\
0.04 0.9986079891689208 \\
0.06 0.9968769429395452 \\
0.08 0.9944638123370607 \\
0.1 0.991374523231391 \\
0.12 0.9876149711942095 \\
0.14 0.9831910164462243 \\
0.16 0.9781084781656838 \\
0.18 0.9723731280665953 \\
0.2 0.9659906831399522 \\
0.22 0.9589667974329371 \\
0.24 0.951307052719155 \\
0.26 0.9430169478866051 \\
0.28 0.934101886838089 \\
0.3 0.9245671646600735 \\
0.32 0.9144179517684831 \\
0.34 0.9036592756818038 \\
0.36 0.8922959999999983 \\
0.38 0.8803328000785169 \\
0.4 0.8677741347752286 \\
0.42 0.8546242135079039 \\
0.44 0.8408869576824477 \\
0.46 0.8265659553255258 \\
0.48 0.811664407464188 \\
0.5 0.7961850644174322 \\
0.52 0.7801301496706918 \\
0.54 0.7635012683514689 \\
0.56 0.7462992964527165 \\
0.58 0.7285242457736957 \\
0.6 0.7101750979388564 \\
0.62 0.6912495986259349 \\
0.64 0.6717439999999982 \\
0.66 0.6516527348723453 \\
0.68 0.6309679995918561 \\
0.7 0.6096792130216032 \\
0.72 0.5877723043219888 \\
0.74 0.5652287595400187 \\
0.76 0.5420243207324984 \\
0.78 0.5181271716008126 \\
0.8 0.4934953415699751 \\
0.82 0.4680728784914328 \\
0.84 0.44178399999999907 \\
0.86 0.41452375807288566 \\
0.88 0.3861423090208506 \\
0.9 0.3564165077437964 \\
0.92 0.32499366214684183 \\
0.94 0.2912647676764824 \\
0.96 0.25401599999999974 \\
0.98 0.21005314276421702 \\
1.0 0.12499999999999979 \\
};
\addlegendentry{UB on $P_{\textnormal{c}}(\code{C}^{(6, 8)}, \set{D}^{\textnormal{ML}})$~\cite[Th.~35]{PolyanskiyPoorVerdu10_1}
}

\addplot [color=mycolor1, dashdotted, mark=triangle, mark options={solid, mycolor1}, each nth point=2, filter discard warning=false]
table[row sep=crcr]{
  0.0 0.9999999873746589 \\
  0.02 0.9996509774410408 \\
  0.04 0.9986079509751824 \\
  0.06 0.9968769429094734 \\
  0.08 0.9944638123248994 \\
  0.1 0.9913745232311951 \\
  0.12 0.9876149711841117 \\
  0.14 0.9831910163837604 \\
  0.16 0.9781084779937338 \\
  0.18 0.9723731277058885 \\
  0.2 0.9659906824863091 \\
  0.22 0.9589667963521286 \\
  0.24 0.9513070510405005 \\
  0.26 0.9430169456264935 \\
  0.28 0.9341018838347147 \\
  0.3 0.9245671606904666 \\
  0.32 0.9144179465457265 \\
  0.34 0.90365926883591 \\
  0.36 0.8922959910461781 \\
  0.38 0.8803327881733779 \\
  0.4 0.8677741193029541 \\
  0.42 0.8546241937994871 \\
  0.44 0.8408869330263506 \\
  0.46 0.8265659249817775 \\
  0.48 0.8116643706816177 \\
  0.5 0.7961850636022171 \\
  0.52 0.7801301487499301 \\
  0.54 0.7635012673029977 \\
  0.56 0.7462992952829967 \\
  0.58 0.728524244492657 \\
  0.6 0.7101750965636844 \\
  0.62 0.6912495971817206 \\
  0.64 0.671743998519569 \\
  0.66 0.6516527333966958 \\
  0.68 0.630967998166861 \\
  0.7 0.6096792116967367 \\
  0.72 0.5877723031460519 \\
  0.74 0.5652287585559089 \\
  0.76 0.542024319972335 \\
  0.78 0.5181271709903282 \\
  0.8 0.4934953411175965 \\
  0.82 0.4680728782485963 \\
  0.84 0.4417839999398432 \\
  0.86 0.41452372388084596 \\
  0.88 0.3861423084733442 \\
  0.9 0.35641650587406426 \\
  0.92 0.32499365754031206 \\
  0.94 0.2912647573493818 \\
  0.96 0.2540159776664126 \\
  0.98 0.2100531242392276 \\
  1.0 0.12499999990538212 \\
};
\addlegendentry{$P_{\textnormal{c}}(\code{C}^{[6, 3]}, \set{D}^{\textnormal{ML}})$}

\addplot [color=blue, dashdotted, mark=asterisk, mark options={solid, blue}, each nth point=2, filter discard warning=false] table[row sep=crcr]{
0.0 0.9999998621753865 \\
0.02 0.9996530528619765 \\
0.04 0.9986248558493795 \\
0.06 0.9969351233481676 \\
0.08 0.9946035653365619 \\
0.1 0.991653065758062 \\
0.12 0.9881033561075371 \\
0.14 0.9839751236017686 \\
0.16 0.9792886851702773 \\
0.18 0.9740649759466227 \\
0.2 0.9683209213799671 \\
0.22 0.9620754431041192 \\
0.24 0.9553434298946651 \\
0.26 0.9481391273880082 \\
0.28 0.9404745908038388 \\
0.3 0.9323598064742975 \\
0.32 0.9238026789534076 \\
0.34 0.9148087543419989 \\
0.36 0.9053815756850243 \\
0.38 0.8955218454286804 \\
0.4 0.8852286205682076 \\
0.42 0.8744987060607223 \\
0.44 0.8633267977403377 \\
0.46 0.8517056962800804 \\
0.48 0.8396263683859069 \\
0.5 0.8270778377058264 \\
0.52 0.8140466201977299 \\
0.54 0.8005194829028821 \\
0.56 0.78647810194997 \\
0.58 0.7719048412203147 \\
0.6 0.7567762658513915 \\
0.62 0.7410682583465441 \\
0.64 0.7247523366649287 \\
0.66 0.7077954128244179 \\
0.68 0.6901593056789121 \\
0.7 0.671799243849652 \\
0.72 0.6526622671611532 \\
0.74 0.6326848233327951 \\
0.76 0.6117894515733688 \\
0.78 0.5898850452001545 \\
0.8 0.5668526945558809 \\
0.82 0.542546948307864 \\
0.84 0.516779547211152 \\
0.86 0.4893026759840584 \\
0.88 0.45978050468854587 \\
0.9 0.42773797035460737 \\
0.92 0.392462373290668 \\
0.94 0.3527914839735794 \\
0.96 0.30654084762554623 \\
0.98 0.24831420975408014 \\
1.0 0.125 \\
};
\addlegendentry{$P_{\textnormal{c}}\bigl(\code{C}^{[6, 3]}, \set{D}^{\star}\bigr)$}

\addplot [color=mycolor2, dashdotted, mark size=2.0pt, mark=o, mark options={solid, mycolor2}, each nth point=2]
  table[row sep=crcr]{%
0	0.943489240454667\\
0.02	0.931379673820261\\
0.04	0.919019249575551\\
0.06	0.906412773836117\\
0.08	0.893565048275628\\
0.1	0.880480868631508\\
0.12	0.867165022997734\\
0.14	0.853622289873433\\
0.16	0.839857435930746\\
0.18	0.825875213459212\\
0.2	0.8116803574365\\
0.22	0.797277582166456\\
0.24	0.782671577414656\\
0.26	0.767867003958782\\
0.28	0.752868488455343\\
0.3	0.737680617505086\\
0.32	0.722307930775934\\
0.34	0.70675491301323\\
0.36	0.691025984731163\\
0.38	0.675125491334418\\
0.4	0.659057690362938\\
0.42	0.642826736481738\\
0.44	0.626436663747639\\
0.46	0.609891364569362\\
0.48	0.59319456462856\\
0.5	0.576349792835637\\
0.52	0.559360345139788\\
0.54	0.542229240675424\\
0.56	0.524959168275261\\
0.58	0.507552420768011\\
0.6	0.490010813638706\\
0.62	0.472335583462002\\
0.64	0.454527259871741\\
0.66	0.436585502469199\\
0.68	0.418508890628373\\
0.7	0.400294649037587\\
0.72	0.381938284029565\\
0.74	0.363433093376056\\
0.76	0.344769493931436\\
0.78	0.325934076888414\\
0.8	0.306908248448369\\
0.82	0.287666213177501\\
0.84	0.268171873799414\\
0.86	0.248373853003327\\
0.88	0.228197055537711\\
0.9	0.207527340493774\\
0.92	0.186180992404673\\
0.94	0.163835507572547\\
0.96	0.139838156771039\\
0.98	0.112445493709397\\
1	0.0625000092899044\\
};
\addlegendentry{LB on $P_{\textnormal{c}}(\code{C}^{(7, 16)}, \set{D}^{\textnormal{ML}})$~\cite[Th.~32]{PolyanskiyPoorVerdu10_1}}

\addplot [color=mycolor2]
  table[row sep=crcr]{%
0	1\\
0.02	0.99947849979594\\
0.04	0.997927999595004\\
0.06	0.995369499332643\\
0.08	0.991823998747488\\
0.1	0.987312497128965\\
0.12	0.98185599293016\\
0.14	0.975475483220362\\
0.16	0.968191962946976\\
0.18	0.960026423970738\\
0.2	0.950999853831326\\
0.22	0.941133234192081\\
0.24	0.930447538902416\\
0.26	0.918963731604141\\
0.28	0.90670276279274\\
0.3	0.893685566226057\\
0.32	0.879933054549743\\
0.34	0.865466113980267\\
0.36	0.850305597850524\\
0.38	0.834472318778256\\
0.4	0.81798703916087\\
0.42	0.800870459628247\\
0.44	0.783143204993085\\
0.46	0.764825807119686\\
0.48	0.745938683978024\\
0.5	0.726502113948299\\
0.52	0.706536204174775\\
0.54	0.68606085141253\\
0.56	0.66509569333221\\
0.58	0.643660047595947\\
0.6	0.6217728351186\\
0.62	0.599452482672396\\
0.64	0.576716798212291\\
0.66	0.553582809734731\\
0.68	0.530066554723387\\
0.7	0.506182801623683\\
0.72	0.481944676212512\\
0.74	0.457363151986474\\
0.76	0.432446343622178\\
0.78	0.40719850342159\\
0.8	0.381618563095608\\
0.82	0.355697949789141\\
0.84	0.329417198371258\\
0.86	0.302740464924214\\
0.88	0.275606151473503\\
0.9	0.24790974334684\\
0.92	0.219469372599922\\
0.94	0.189947190858707\\
0.96	0.158630396465842\\
0.98	0.123555310430844\\
1	0.062500010860192\\
};
\addlegendentry{UB on $P_{\textnormal{c}}(\code{C}^{(7, 16)}, \set{D}^{\textnormal{ML}})$~\cite[Th.~35]{PolyanskiyPoorVerdu10_1}}

\addplot [color=mycolor2, dashdotted, mark=triangle, mark options={solid, mycolor2}, each nth point=2]
  table[row sep=crcr]{%
0	1\\
0.02	0.99947849979594\\
0.04	0.997927999595004\\
0.06	0.995369499332643\\
0.08	0.991823998747488\\
0.1	0.987312497128966\\
0.12	0.98185599293016\\
0.14	0.975475483220362\\
0.16	0.968191962946977\\
0.18	0.960026423970738\\
0.2	0.950999853831326\\
0.22	0.941133234192081\\
0.24	0.930447538902416\\
0.26	0.918963731604141\\
0.28	0.90670276279274\\
0.3	0.893685566226057\\
0.32	0.879933054549743\\
0.34	0.865466113980267\\
0.36	0.850305597850524\\
0.38	0.834472318778256\\
0.4	0.81798703916087\\
0.42	0.800870459628247\\
0.44	0.783143204993085\\
0.46	0.764825807119686\\
0.48	0.745938683978024\\
0.5	0.726502113948299\\
0.52	0.706536204174775\\
0.54	0.686060851412529\\
0.56	0.66509569333221\\
0.58	0.643660047595947\\
0.6	0.6217728351186\\
0.62	0.599452482672396\\
0.64	0.576716798212291\\
0.66	0.553582809734731\\
0.68	0.530066554723387\\
0.7	0.506182801623683\\
0.72	0.481944676212512\\
0.74	0.457363151986474\\
0.76	0.432446343622178\\
0.78	0.40719850342159\\
0.8	0.381618563095608\\
0.82	0.355697949789141\\
0.84	0.329417198371258\\
0.86	0.302740464924213\\
0.88	0.275606151473504\\
0.9	0.247909743346841\\
0.92	0.219469372599922\\
0.94	0.189947190858707\\
0.96	0.158630396465842\\
0.98	0.123555310430844\\
1	0.0625000108601922\\
};
\addlegendentry{$P_{\textnormal{c}}(\code{C}^{[7, 4]}, \set{D}^{\textnormal{ML}})$}

\addplot [color=red, dashdotted, mark=asterisk, mark options={solid, red}, each nth point=2]
  table[row sep=crcr]{%
0	0.999999999999428\\
0.02	0.999478703724007\\
0.04	0.997931160173189\\
0.06	0.995384999279701\\
0.08	0.991871421387987\\
0.1	0.98742448831235\\
0.12	0.982080399115277\\
0.14	0.975876875738362\\
0.16	0.96885267461903\\
0.18	0.961046308606224\\
0.2	0.952496240340801\\
0.22	0.943239897189549\\
0.24	0.933313259801638\\
0.26	0.92275045330065\\
0.28	0.911583372486257\\
0.3	0.899841441019868\\
0.32	0.887551331983246\\
0.34	0.874737472866618\\
0.36	0.861420697580632\\
0.38	0.847619677521012\\
0.4	0.833350015606676\\
0.42	0.818624740521786\\
0.44	0.803454150152493\\
0.46	0.787845806352005\\
0.48	0.771804422266418\\
0.5	0.755332725459817\\
0.52	0.738429202988733\\
0.54	0.721090206534724\\
0.56	0.703308650870947\\
0.58	0.685073978808427\\
0.6	0.666371905108908\\
0.62	0.647184214204386\\
0.64	0.627488278633851\\
0.66	0.607256854728827\\
0.68	0.586457364717725\\
0.7	0.565052327670781\\
0.72	0.542996858177913\\
0.74	0.520239526173973\\
0.76	0.496721329969085\\
0.78	0.472373019820033\\
0.8	0.447113317673203\\
0.82	0.420846897092814\\
0.84	0.39345841059999\\
0.86	0.364805387278695\\
0.88	0.33470522289613\\
0.9	0.302911162957837\\
0.92	0.269065216420989\\
0.94	0.232591593280953\\
0.96	0.192396307808215\\
0.98	0.145683734506476\\
1	0.0625000000173412\\
};
\addlegendentry{$P_{\textnormal{c}}\bigl(\code{C}^{[7, 4]}, \set{D}^{\star}\bigr)$}

\end{axis}
\end{tikzpicture}

%% file: figs/ADC-success_prob_vs_gamma_n6_k4_q2.tex



\definecolor{mycolor1}{rgb}{0.30100,0.74500,0.93300}

\begin{tikzpicture}

\begin{axis}[
width=7.75cm, height=6.50cm, at={(1.011in,0.642in)},
scale only axis,
xmin=0, xmax=1, xlabel style={font=\color{white!15!black}}, xlabel={The damping probability $\gamma$},
ymin=0.0625, ymax=1.00, yminorticks=true, ylabel style={font=\color{white!15!black}}, 
ylabel={$P_{\textnormal{c}}$}, ylabel style = {yshift=-1mm}, 
axis background/.style={fill=white},
xmajorgrids, ymajorgrids, 
legend style={at={(0.00,0.00)}, anchor=south west, legend cell align=left, align=left, draw=white!15!black}, font=\scriptsize
]

		\addplot [color=mycolor1, dashdotted, mark size=2.0pt, mark=o, mark options={solid, mycolor1}, mark repeat=2] table[row sep=crcr]{
			0.0 0.8909392960613602 \\
			0.05 0.8549487422220355 \\
			0.1 0.8185781345071323 \\
			0.15 0.781849510990669 \\
			0.2 0.7447841444710717 \\
			0.25 0.7074022737663697 \\
			0.3 0.6697227091976756 \\
			0.35 0.6317622838549409 \\
			0.4 0.5935350821425263 \\
			0.45 0.5550513306098412 \\
			0.5 0.5163157917885237 \\
			0.55 0.47732517272044017 \\
			0.6 0.4380643821514061 \\
			0.65 0.3985000733709419 \\
			0.7 0.35856967497339404 \\
			0.75 0.3181610885269708 \\
			0.8 0.27707159808361626 \\
			0.85 0.23491326531210946 \\
			0.9 0.1908474470634806 \\
			0.95 0.14251396044059625 \\
			1.0 0.06249999995269105 \\
		};
		\addlegendentry{LB on $P_{\textnormal{c}}(\code{C}^{[6, 4]_2}, \set{D}^{\textnormal{ML}})$~\cite[Th.~32]{PolyanskiyPoorVerdu10_1}}

		\addplot [color=mycolor1] table[row sep=crcr]{
			0.0 0.9999999780328638 \\
			0.05 0.962039061498279 \\
			0.1 0.9231874672076594 \\
			0.15 0.8834919229949872 \\
			0.2 0.8429988131398174 \\
			0.25 0.8017539448356334 \\
			0.3 0.7598021896189254 \\
			0.35 0.7171869722567459 \\
			0.4 0.6739495375127078 \\
			0.45 0.6301278765346562 \\
			0.5 0.5857551510223347 \\
			0.55 0.5408570904907117 \\
			0.6 0.4954482259750978 \\
			0.65 0.44952527390559227 \\
			0.7 0.40305579910761413 \\
			0.75 0.35595703067727574 \\
			0.8 0.30805262216821383 \\
			0.85 0.25897232662771863 \\
			0.9 0.20786846217333965 \\
			0.95 0.15226444471772482 \\
			1.0 0.06249999995269105 \\
		};
		\addlegendentry{UB on $P_{\textnormal{c}}(\code{C}^{[6, 4]_2}, \set{D}^{\textnormal{ML}})$~\cite[Th.~35]{PolyanskiyPoorVerdu10_1}}

		\addplot [color=mycolor1, dashdotted, mark=triangle, mark options={solid, mycolor1}, each nth point=2, filter discard warning=false] table[row sep=crcr]{
			0.0 0.9999999999999969 \\
			0.05 0.9503124743094337 \\
			0.1 0.9012495665739917 \\
			0.15 0.8528101811497938 \\
			0.2 0.8049922359499591 \\
			0.25 0.7577923641556891 \\
			0.3 0.7112055112769812 \\
			0.35 0.665224382117313 \\
			0.4 0.6198386676965907 \\
			0.45 0.5750339413749558 \\
			0.5 0.5307900429449542 \\
			0.55 0.48707864255310124 \\
			0.6 0.44385943621178525 \\
			0.65 0.40107394267961083 \\
			0.7 0.3586348311891782 \\
			0.75 0.31640624999999944 \\
			0.8 0.2741640786499865 \\
			0.85 0.2315045212034622 \\
			0.9 0.18758763565462974 \\
			0.95 0.1401030344093691 \\
			1.0 0.062499999999999896 \\
		};
		\addlegendentry{$P_{\textnormal{c}}(\code{C}^{[6, 4]_2}, \set{D}^{\textnormal{ML}})$}

            \addplot [color=blue, dashdotted, mark=asterisk, mark options={solid, blue}, each nth point=2, filter discard warning=false] table[row sep=crcr]{
			0.0 0.9999999999819463 \\
			0.05 0.9526497824561714 \\
			0.1 0.909907857925758 \\
			0.15 0.870689650493314 \\
			0.2 0.8339243726275494 \\
			0.25 0.7986282507378644 \\
			0.3 0.7639499404690533 \\
			0.35 0.7291784458811306 \\
			0.4 0.693742755127704 \\
			0.45 0.6571912982066868 \\
			0.5 0.6191748036418683 \\
			0.55 0.579421131496713 \\
			0.6 0.5377134519481807 \\
			0.65 0.49386487021911585 \\
			0.7 0.44769428338639033 \\
			0.75 0.39899294170723787 \\
			0.8 0.3474749209835266 \\
			0.85 0.29267063852151726 \\
			0.9 0.23363259763911196 \\
			0.95 0.16772147781860955 \\
			1.0 0.06249999999999913 \\
		};
		\addlegendentry{$P_{\textnormal{c}}(\code{C}^{[6, 4]_2}, \set{D}^*)$}

  \end{axis}
\end{tikzpicture}

%% file: figs/EA_ADC-success_prob_vs_gamma_n3_k1_q4.tex



    \definecolor{mycolor1}{rgb}{0.30100,0.74500,0.93300}

    \begin{tikzpicture}
        \begin{axis}[
            width=7.75cm, height=6.50cm, at={(1.011in,0.642in)},
            scale only axis,
            xmin=0, xmax=1, xlabel style={font=\color{white!15!black}}, xlabel={The damping probability $\gamma$},
            ymin=0.25, ymax=1.00, yminorticks=true, ylabel style={font=\color{white!15!black}}, 
            ylabel={$P_{\textnormal{c}}$}, ylabel style = {yshift=-1mm}, 
            axis background/.style={fill=white},
            xmajorgrids, ymajorgrids, 
            legend style={at={(0.00,0.00)}, anchor=south west, legend cell align=left, align=left, draw=white!15!black}, font=\scriptsize
        ]
		\addplot [color=blue, dashdotted, mark=asterisk, mark options={solid, blue}, each nth point=2, filter discard warning=false] table[row sep=crcr]{
			0.0 1.000000000001699 \\
			0.02 0.9999170919543011 \\
			0.04 0.9996623311573791 \\
			0.06 0.9992268653980636 \\
			0.08 0.9986004586305139 \\
			0.1 0.9977743244712354 \\
			0.12 0.9967371965676523 \\
			0.14 0.9954786216319804 \\
			0.16 0.9939873928929384 \\
			0.18 0.9922518674307976 \\
			0.2 0.9902598442664843 \\
			0.22 0.9879985149632318 \\
			0.24 0.9854544314225131 \\
			0.26 0.9826137181354999 \\
			0.28 0.9794614593649182 \\
			0.3 0.9759821683549467 \\
			0.32 0.9721595289747932 \\
			0.34 0.9679762593253943 \\
			0.36 0.9634141553021326 \\
			0.38 0.9584539470203288 \\
			0.4 0.9530751760421103 \\
			0.42 0.9472563238131098 \\
			0.44 0.9409743117661313 \\
			0.46 0.9342046431354771 \\
			0.48 0.926921202080986 \\
			0.5 0.9190962558445458 \\
			0.52 0.9106998566452895 \\
			0.54 0.9017000399848502 \\
			0.56 0.8920624039150957 \\
			0.58 0.8817497951099555 \\
			0.6 0.8707218730694695 \\
			0.62 0.8589352831998922 \\
			0.64 0.8463419133651571 \\
			0.66 0.8328893208940427 \\
			0.68 0.8185194978212336 \\
			0.7 0.8031677210279868 \\
			0.72 0.7867611267937177 \\
			0.74 0.7692169195883869 \\
			0.76 0.750439733418913 \\
			0.78 0.7303181068859009 \\
			0.8 0.7087194099551581 \\
			0.82 0.6854825669041275 \\
			0.84 0.660406862318726 \\
			0.86 0.633234721560612 \\
			0.88 0.6036206052097768 \\
			0.9 0.5710809075752356 \\
			0.92 0.5348895469727768 \\
			0.94 0.49384784115788993 \\
			0.96 0.44565884276850853 \\
			0.98 0.38448265210415145 \\
			1.0 0.24999999999809874 \\
		};
		\addlegendentry{$P_{\textnormal{c}}(\code{C}^{[3, 1]_4}, \set{D}^*)$}

		\addplot [color=mycolor1] table[row sep=crcr]{
			0.0 0.9999999969796812 \\
			0.02 0.999848990295417 \\
			0.04 0.9993919024192954 \\
			0.06 0.998622524785749 \\
			0.08 0.9975344776947934 \\
			0.1 0.9961212305445857 \\
			0.12 0.9943760694208094 \\
			0.14 0.9922920882194564 \\
			0.16 0.9898621736197561 \\
			0.18 0.9870789884759711 \\
			0.2 0.9839349534138888 \\
			0.22 0.9804222263983653 \\
			0.24 0.9765326799853636 \\
			0.26 0.9722578759264516 \\
			0.28 0.967589036735439 \\
			0.3 0.9625170137568193 \\
			0.32 0.957032251190233 \\
			0.34 0.9511247353087722 \\
			0.36 0.9447839895178163 \\
			0.38 0.9379989589169822 \\
			0.4 0.9307579928266497 \\
			0.42 0.9230487650070034 \\
			0.44 0.9148581940740743 \\
			0.46 0.9061723512430762 \\
			0.48 0.89697635657794 \\
			0.5 0.8872542364096969 \\
			0.52 0.8769888022013284 \\
			0.54 0.8661614537486693 \\
			0.56 0.8547519707262043 \\
			0.58 0.8427382692884093 \\
			0.6 0.8300960856552937 \\
			0.62 0.8167986015313855 \\
			0.64 0.8028159775739716 \\
			0.66 0.7881147725714772 \\
			0.68 0.7726571938808324 \\
			0.7 0.7564001690920423 \\
			0.72 0.7392941776288806 \\
			0.74 0.7212813357079211 \\
			0.76 0.7022935831089268 \\
			0.78 0.6822493933301255 \\
			0.8 0.6610495140507944 \\
			0.82 0.6385706644864309 \\
			0.84 0.614655994424658 \\
			0.86 0.5890999953392956 \\
			0.88 0.5616233055070412 \\
			0.9 0.5318276163073415 \\
			0.92 0.49910715879386913 \\
			0.94 0.462450956211321 \\
			0.96 0.4199039320422074 \\
			0.98 0.3664594103443282 \\
			1.0 0.2499999999975121 \\
		};
		\addlegendentry{UB on $P_{\textnormal{c}}(\code{C}^{[3, 1]_4}, \set{D}^{(\textnormal{ML})})$ \cref{thm:UB_qAC}}

		\addplot [color=mycolor1, dashdotted, mark=triangle, mark options={solid, mycolor1}, each nth point=2, filter discard warning=false] table[row sep=crcr]{
			0.0 0.9999999999999987 \\
			0.02 0.9998489942988885 \\
			0.04 0.9993919075277665 \\
			0.06 0.9986225253138801 \\
			0.08 0.997534478424522 \\
			0.1 0.9961212314765182 \\
			0.12 0.9943760705412825 \\
			0.14 0.9922920895090199 \\
			0.16 0.9898621750552283 \\
			0.18 0.9870789900284874 \\
			0.2 0.9839349550499532 \\
			0.22 0.9804222280811944 \\
			0.24 0.9765326816765544 \\
			0.26 0.9722578775879073 \\
			0.28 0.9675890383315967 \\
			0.3 0.9625170152570917 \\
			0.32 0.9570322525716473 \\
			0.34 0.951124746671202 \\
			0.36 0.944783999999999 \\
			0.38 0.9379989685037011 \\
			0.4 0.9307580015448891 \\
			0.42 0.9230487729047658 \\
			0.44 0.9148582011864204 \\
			0.46 0.9061723575434959 \\
			0.48 0.8969763581575336 \\
			0.5 0.8872542382415916 \\
			0.52 0.8769888035073559 \\
			0.54 0.8661614539284542 \\
			0.56 0.8547519731654901 \\
			0.58 0.8427382750473063 \\
			0.6 0.8300960958218888 \\
			0.62 0.8167986171944666 \\
			0.64 0.8028159999999983 \\
			0.66 0.788114801007211 \\
			0.68 0.7726572347159889 \\
			0.7 0.7564002263159856 \\
			0.72 0.7392941783047838 \\
			0.74 0.7212813366811055 \\
			0.76 0.7022935845040782 \\
			0.78 0.682249395307727 \\
			0.8 0.6610495168499689 \\
			0.82 0.6385706684419086 \\
			0.84 0.6146559999999988 \\
			0.86 0.5891000031543364 \\
			0.88 0.5616233162899175 \\
			0.9 0.5318276303622073 \\
			0.92 0.4991071801064709 \\
			0.94 0.4624509960945625 \\
			0.96 0.41990399999999917 \\
			0.98 0.3664594103967953 \\
			1.0 0.2499999999999999 \\
		};
            \addlegendentry{$P_{\textnormal{c}}(\code{C}^{[3, 1]_4}, \set{D}^{(\textnormal{ML})})$}

		\addplot [color=mycolor1, dashed, mark size=2.0pt, mark=o, mark options={solid, mycolor1}, mark repeat=2] table[row sep=crcr]{
			0.0 0.9768056763894588 \\
			0.02 0.9721837457088567 \\
			0.04 0.9673624243060254 \\
			0.06 0.962338941228127 \\
			0.08 0.9571103632508081 \\
			0.1 0.9516736717731976 \\
			0.12 0.9460256974597376 \\
			0.14 0.9401631230895117 \\
			0.16 0.9340824717800339 \\
			0.18 0.9277800939569911 \\
			0.2 0.9212521528981243 \\
			0.22 0.91449460866855 \\
			0.24 0.907503200217594 \\
			0.26 0.9002734253718339 \\
			0.28 0.8928005184122586 \\
			0.3 0.8850794248672331 \\
			0.32 0.877104773084272 \\
			0.34 0.8688708295654707 \\
			0.36 0.8603715136876797 \\
			0.38 0.8516002771969826 \\
			0.4 0.8425501104365192 \\
			0.42 0.8332134737669259 \\
			0.44 0.8235822342914576 \\
			0.46 0.8136475924282507 \\
			0.48 0.8034000002375534 \\
			0.5 0.7928290432981251 \\
			0.52 0.781923351312078 \\
			0.54 0.7706704353023957 \\
			0.56 0.7590565218094902 \\
			0.58 0.7470663597604568 \\
			0.6 0.7346829651556603 \\
			0.62 0.7218873208806982 \\
			0.64 0.7086580018447641 \\
			0.66 0.694970708082557 \\
			0.68 0.6807976600313079 \\
			0.7 0.6661068523069344 \\
			0.72 0.6508611153291197 \\
			0.74 0.6350165468037827 \\
			0.76 0.6185210764091167 \\
			0.78 0.6013117831735881 \\
			0.8 0.5833114480523792 \\
			0.82 0.5644234405614345 \\
			0.84 0.5445239695846478 \\
			0.86 0.5234498145075785 \\
			0.88 0.5009778161223158 \\
			0.9 0.47678812794318465 \\
			0.92 0.4503920001842167 \\
			0.94 0.4209702529780708 \\
			0.96 0.3869325561958389 \\
			0.98 0.34419159368676194 \\
			1.0 0.24999999999751207 \\
		};
		\addlegendentry{LB on $P_{\textnormal{c}}(\code{C}^{[3, 1]_4}, \set{D}^{(\textnormal{ML})})$  \cref{thm:LB_qAC}}
        
        \end{axis}
    \end{tikzpicture}